\documentclass[%
 reprint,
 amsmath,amssymb,
prstper,
]{revtex4-2}

\usepackage{graphicx}
\usepackage{dcolumn}
\usepackage{bm}

\usepackage{hyperref}
\hypersetup{
    colorlinks = true,
    linkcolor = blue,
    anchorcolor = blue,
    citecolor = blue,
    filecolor = blue,
    urlcolor = blue
    }
\usepackage[super]{nth}
\usepackage{wasysym}
\usepackage{booktabs}

\newenvironment{myindent}
{\par\leftskip0.5cm\relax\rightskip0.5cm\relax}
{\par\leftskip0cm\relax\rightskip0cm\relax}

\begin{document}

\preprint{APS/123-QED}

\title{Nuclear History, Politics, and Futures from (A)toms-to(Z)oom: \\ Design and Deployment of a Remote-Learning Special-Topics  \\ Course For Nuclear Engineering Education}%

\author{Aaron J. Berliner}
\author{Jake Hecla}
\affiliation{%
Department of Nuclear Engineering, University of California Berkeley, Berkeley, CA}%




\date{\today}

\begin{abstract}

To address the lack of familiarity with nuclear history common among nuclear engineers and physicists, we outline the design and deployment of a special-topics course entitled ``NE290: Nuclear History, Politics, and Futures'' across which we contextualize the importance of the field at its inception, in current affairs, and in future endeavors. We argue that understanding this history is paramount in internalizing a sense of respect for the scientific, technical, and sociological ramifications of an unlocked atom -- as well as their perils. We begin by outlining the gaps in secondary educational offerings for nuclear history and their importance in consideration with nontechnical engineering guidelines. We then outline a number ABET specifications as pedagogical goals for NE290 from which we derive a list of target student learning objectives. Next, we outline the NE290 syllabus in terms of assignments and an overview of course content in the form of a class timeline. We provide an extensive description of the materials and teaching methodologies for the four units of NE290: 20th-Century Physics, Physics in WWII, the Early Cold War, and the Late Cold War and Modern Era. We detail the sequence of lectures across the course and historical timelines -- leading up to a showcasing of NE290 final projects which mirror in creativity the novelty of course offering. Because NE290 was first offered during Spring 2021 during the COVID-19 pandemic, additional measures in the form of new tools were used to augment the mandate of remote learning. In particular, we leveraged the newfound ubiquity of videoconferencing technology to recruit geographically diverse guest lecturers, and used the MIRO tool for virtual whiteboarding. Lastly, we provide an accounting of course outcomes drawn from student feedback which – in tandem with the complete distribution of course material – facilitates the integration of nuclear history into the curriculum for the wider nuclear engineering and physics communities.

\end{abstract}

\maketitle


\section{Introduction}

As of 2021, the Accreditation Board for Engineering and Technology (ABET) program criteria $(\mathfrak{C})$ for Nuclear, Radiological, and Similarly Named Engineering programs require curriculum\cite{ABET2021Criteria2020-2021} in
\begin{itemize}
    \item[$\mathfrak{C}_{1}$.] mathematics, to support analyses of complex nuclear or radiological problems; 
    \item[$\mathfrak{C}_{2}$.] atomic and nuclear physics;
    \item[$\mathfrak{C}_{3}$.] transport and interaction of radiation with matter; 
    \item[$\mathfrak{C}_{4}$.] nuclear or radiological systems and processes; 
    \item[$\mathfrak{C}_{5}$.] nuclear fuel cycles; 
    \item[$\mathfrak{C}_{6}$.] nuclear radiation detection and measurement; and 
    \item[$\mathfrak{C}_{7}$.] nuclear or radiological system design
\end{itemize}
While a mastery of the material presented in the summary of the general criteria for Master’s level and integrated Baccalaureate and Master’s level engineering programs would enable sufficient depth for engineering practice, there is no requirement of any criterion that would aid in a student's understanding of the history of nuclear energy and its interaction with society. For instance, the aforementioned ABET technical criteria would not aide in understanding
\begin{itemize}
    \item[$\mathfrak{H}_{1}$.] how Henri Poincare's (1854-1912)'s mathematical discovery in 1905 was delayed in publication until a 1906 paper\cite{Poincare1906SurLelectron} in which he extended the work of Hendrik Lorentz (1853-1928) to show radiation could be considered ``a fictitious fluid'' with an equivalent mass of $E=mc^{2}$, preempting Einstein's (1879-1955) 1905 work\cite{Einstein1905ZurKorper} on special relativity. 
    \item[$\mathfrak{H}_{2}$.] how atomic and nuclear physics of the early \nth{20}-century was shaped by the efforts in quantum mechanics enabled by Lorentz's 1903 meeting of Paul Ehrenfest\cite{Huijnen2007Paul19041912} (1880-1933) and their collaboration with Einstein\cite{Illy1989Einstein19131920}.
    \item[$\mathfrak{H}_{3}$.] how the transport and interaction of radiation with matter could be understood through the equations of Ehrenfest's student Gregory Breit\cite{Ehrenfest1922AQuantization} (1899-1981) in collaboration with Eugene Wigner\cite{Breit1936CaptureNeutrons} (1902-1995).
    \item[$\mathfrak{H}_{4}$.] how Wigner's efforts in the Manhattan project aided a fellow Ehrenfest pupil, Robert Oppenheimer (1904-1967) to weaponize the first nuclear system; 
    \item[$\mathfrak{H}_{5}$.] how Oppenheimer's efforts at Los Alamos were aided by those at the Met-Lab of another Ehrenfest student, Enrico Fermi (1901-1954) to leverage nuclear fuel cycles in the production of fissile materials\cite{Enrico1955NeutronicReactor,Enrico1940ProcessSubstances}; 
    \item[$\mathfrak{H}_{6}$.] how nuclear radiation detection and measurement of the $\beta$ spectra was enabled through the 1939 efforts of Robert Oppenheimer's brother, Frank Oppenheimer\cite{Oppenheimer1939BetaSpectra} (1912-1985).
    \item[$\mathfrak{H}_{7}$.] how the methods of nuclear or radiological system design today are heavily influenced by the efforts of Fermi in Chicago\cite{Donnell1958EnricoPlant,Hargraves2010LiquidReexamined} and many more mathematicians, scientists, and engineers whose names should not be forgotten.
\end{itemize}
While the ``hows'' described above ($\mathfrak{H}$) do not factor into the technical aspects ($\mathfrak{H}$) required for the engineering challenges that await B.S. and M.S. graduates of ABET certified programs, the importance of understanding the historical context as described above enriches the field\cite{Olesko2006ScienceFuture,Monk1997PlacingPedagogy}. For aspiring students still in secondary education, the framing of Nuclear Engineering in terms of an evolving story in shades of 6-degrees-of-separation offer an alternative pedagogical pathway to learning\cite{Foster2010PublicCases}. In undergraduate and graduate STEM programs, such anecdotes are most often found in a slide-or-two as preface for a technical discussion of engineering principles. For students at the undergraduate and graduate level, ABET has provided a Code of Ethics of Engineers replete with a number of fundamental principles and canons\cite{10.2307/1014589} to guide future engineer as they uphold and advance the integrity, honor and dignity of the engineering profession. However, there can be no advance of engineering integrity, honor, and dignity without an appreciation of the history that shaped the transformation from ideas and imagination to realized engineering marvels. 

To address the lack of familiarity with nuclear history common among nuclear engineers and physicists, we designed a special-topics course entitled ``NE290: Nuclear History, Politics, and Futures'' across which we contextualize the importance of the field at its inception, in current affairs, and in future endeavors. We argue that understanding this history is paramount in internalizing a sense of respect for the fruits of an unlocked atom, as well as its perils. 

\section{NE290 Description}
The NE290 course was designed to address and exceed the nontechnical ABET specifications for Student Outcomes ($\mathfrak{O}$), primarily:
\begin{enumerate}
    \item[$\mathfrak{O}_{1}$.] an ability to recognize ethical and professional responsibilities in engineering situations and make informed judgments, which must consider the impact of engineering solutions in global, economic, environmental, and societal contexts.
    \item[$\mathfrak{O}_{2}$.] an ability to function effectively on a team whose members together provide leadership, create a collaborative and inclusive environment, establish goals, plan tasks, and meet objectives.
    \item[$\mathfrak{O}_{3}$.] an ability to acquire and apply new knowledge as needed, using appropriate learning strategies.
\end{enumerate}
These nontechnical outcomes correspond to many of the precepts that would be gained through the proposed formal historical education.

\begin{figure*}
    \includegraphics[width=17.2cm]{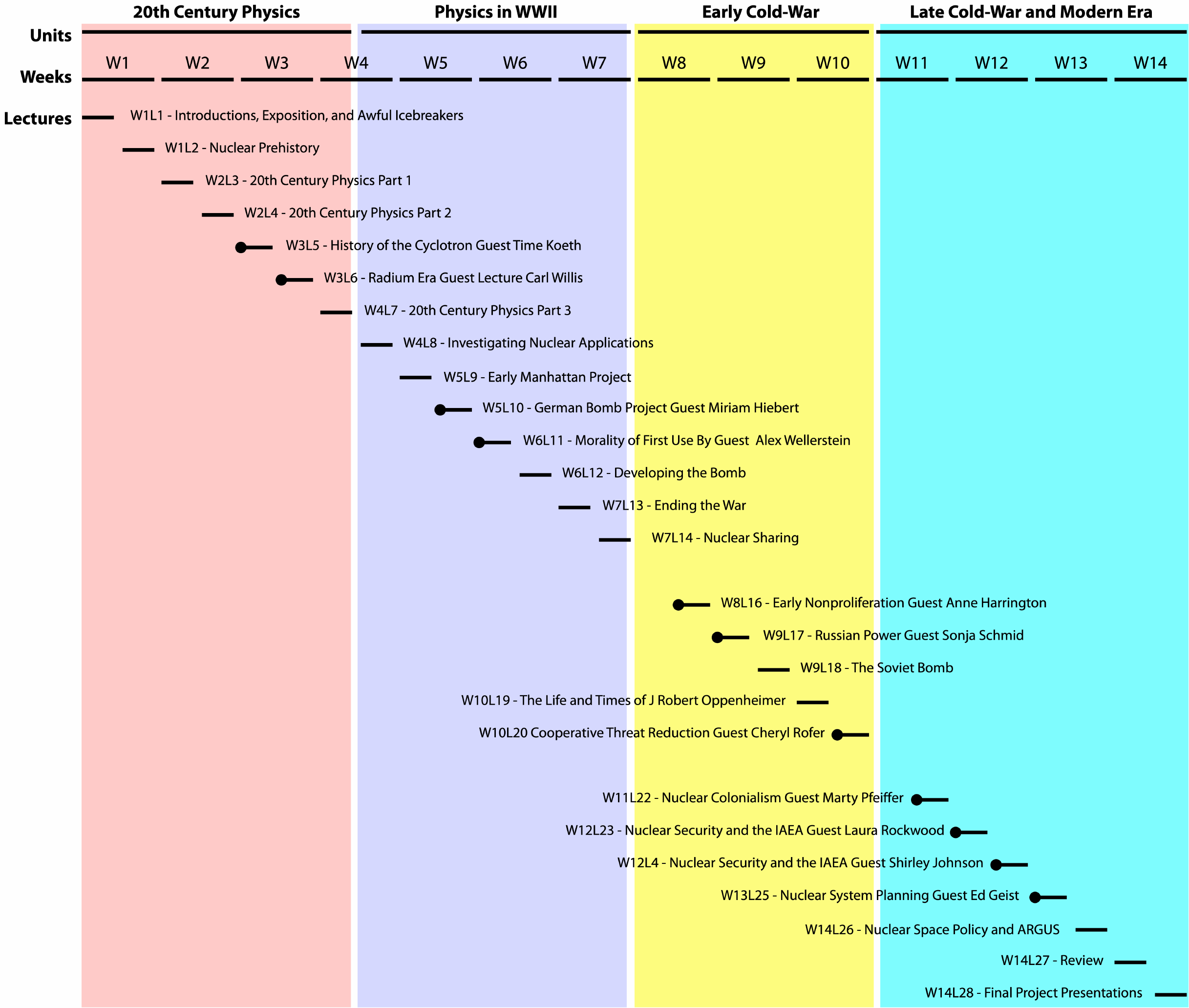}
    \caption{Graphical syllabus for NE290 course as broken down hierarchically in terms of units, weeks, and lectures. Units are colored. Guest lectures are indicated by a $\fullmoon$ marker. All lectures slides are provided as Supporting Information.}
    \label{fig:syllabus} 
\end{figure*}

NE290 spans over a century of nuclear history. We began with a unit on \nth{20}-century developments in fundamental physics and mathematics that evolved alongside the first experimental evidence of atomic and nuclear structure. Our next unit described the lead-up to and developments of the Manhattan project, as well as its foreign counterparts. We then explored the early atomic age with a look at how the growing tension with the Soviets led to an arms race that dominated foreign policy for decades. This unit offered additional focus on the era of nuclear-adjacent technologies such as strategic bombers, as well as the development of the nuclear submarine, the space race, and the hydrogen bomb. In along this timeline of key events, we explored the social and political aspects of the field through literature that speaks to the tolls of the nuclear complex, nuclear testing, and the growing disillusionment and terror inspired by nuclear technology. We also explored the still-present shadows of nuclear winter and the evolution of post-Cold War nuclear arsenals. Throughout each of these units, we organized the lectures and assignments in accordance with following set of learning outcomes ($\mathfrak{B}$) based on Bloom's Taxonomy\cite{Bloom1984BloomObjectives}
\begin{enumerate}
    \item[$\mathfrak{B}_{1}$] \textbf{Analyze} how the Manhattan project was influenced by these discoveries.
    \item[$\mathfrak{B}_{2}$] \textbf{Sequence} the complex historical basis for nuclear armament.
    \item[$\mathfrak{B}_{3}$] \textbf{Summarize} the landmark players that shaped the nuclear engineering communities.
    \item[$\mathfrak{B}_{4}$] \textbf{Summarize} the persistent problems in nuclear policy and engineering from a historical perspective.
    \item[$\mathfrak{B}_{5}$] \textbf{Analyze} how current persistent problems in nuclear policy and engineering can be related to problems solved through a historical perspective.
    \item[$\mathfrak{B}_{6}$] \textbf{Synthesize} solutions to current persistent problems in nuclear policy and engineering from solutions to past problems.
    \item[$\mathfrak{B}_{7}$] \textbf{Analyze} impact of nuclear physics on international relations and world affairs.
\end{enumerate}
These learning outcomes roughly correspond to the the breakdown of the NE290 timeline of nuclear history into units as shown in Figure \ref{fig:syllabus}. Throughout the semester, the class was graded based on the following:
\begin{itemize}
    \item Weekly Reading Responses 30\%
    \item Class Participation 20\%
    \item Term Paper 50\%
\end{itemize}

\section{Active Reading Assignments}
Much of the process for becoming conversant in history is active reading. We prepared a schedule with a wide array of readings spanning historical biographies to social science literature. Whenever possible, we also provided media in terms of audiobooks, films, and artwork to augment the learning process. 
Students were expected to provide a thoughtful weekly response of $\sim$1 page to the reading materials and class lectures. Each week, a random selection of students where be asked to share their responses with the class to foster a discussion, so students will need to be prepared to engage and discuss both the literature and their interpretation of it. Becoming conversant in history means developing a faculty for creatively processing the past in the present for a better future, and to aid in this, we prepared an interactive MIRO board across which we will all be posting materials and adding comments and suggestions. When not asked for a response page, students were assigned with the task to add content to MIRO that they felt brings the history to life.

\section{Course Content}
\subsection{\nth{20}-Century Physics}
NE290 began with a unit on the history of physics from antiquity through \nth{20}-Century physics. Our lecture on nuclear prehistory (W1L2) focused on natural nuclear reactors\cite{Mathieu2001AlterationGabon,Ebisuzaki2017NuclearLife,Jensen1996Uraninite:Africa}, first encounters with radiation-induced illness\cite{Robison2006St.Cancer}, and the initial industrial uses of Uranium\cite{Caley1948TheUranium}. This lecture was intended to give broad background on natural radioactivity and non-nuclear uses of nuclear material. 


\begin{figure*}[t!]
    \centering
    \includegraphics[width=17.2cm]{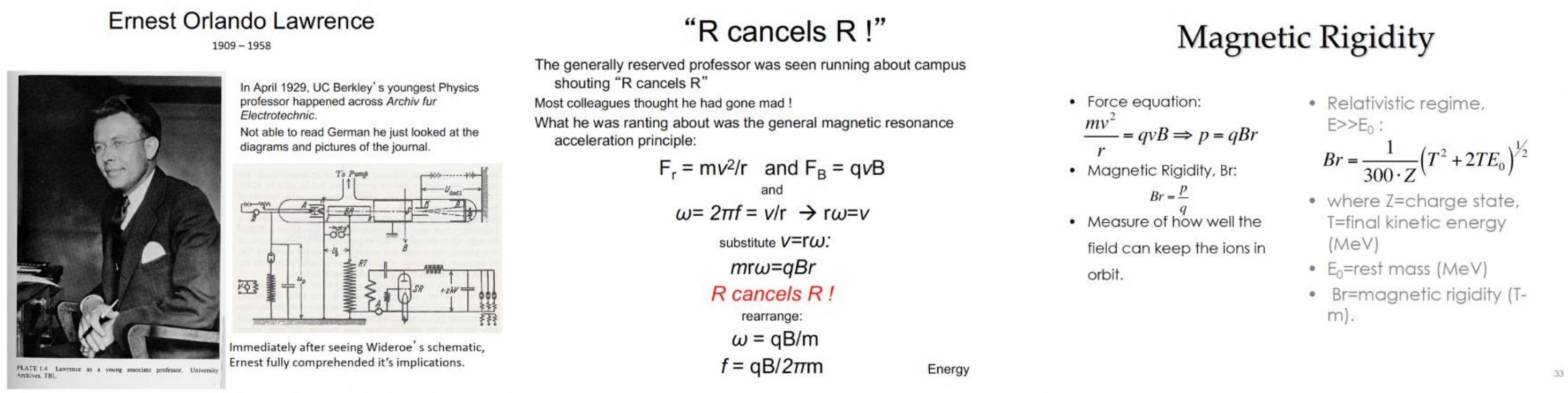}
    \caption{Cyclotron History Slides. Lecture W3L5.}
    \label{fig:cyclo1}
\end{figure*}

\begin{figure}[h!]
    \centering
    4\includegraphics[width=8.6cm]{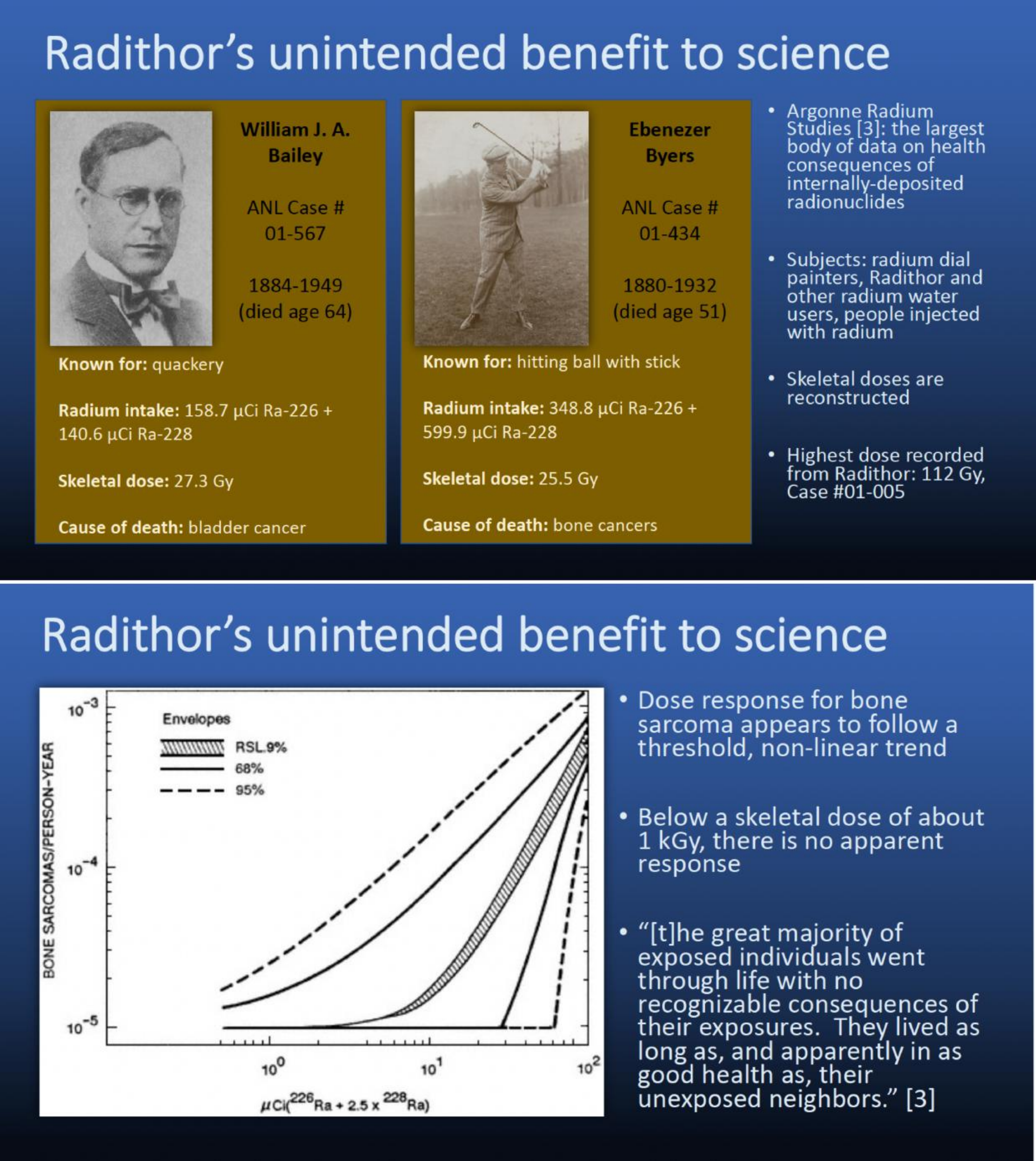}
    \caption{Radium Era Slides. Lecture W3L6.}
    \label{fig:quakery1}
\end{figure}

We began our initial history lecture with a discussion of some of the first observations of plasma effects such as the 1675 ``ghostly lights in barometers''\cite{Banks2009Starting1665-1700} and the 1719 developments in the ``influence matchine'' that could produce significant Mercury discharge and its relationship to ``exceed the performance of cat fur and a glass rod\cite{Picard1676.Barometre}. This later bit of humor set the tone for the course in terms of our use of humor and anecdotes while also playing a part of the basis for the 1880s development of cathode rays\cite{Braun1897UberStrome}, 1985 discovery of x-rays by Rontgen\cite{Rontgen1895OnCommunication}, 1896 discovery of radioactivity by Becquerel\cite{Becquerel1896SurPhosphorescence}, 1897 discovery of the electron by Thomson\cite{Thomson1897CathodeIn}, and the 1890s efforts by Curie leading to the co-discovery of Radium\cite{Curie1898RayonsThorium}. Lectures by the primary teaching team were then augmented by guest lectures on the history of the cyclotron (W3L5) by Dr. Tim Koeth (University of Maryland, Figure \ref{fig:cyclo1}) the ``Radium Era'' (W3L6, Figure \ref{fig:quakery1}) by Dr. Carl Willis (University of New Mexico) which combined early nuclear history with technical elements of engineering. 

\begin{figure}[h!]
    \centering
    \includegraphics[width=8.6cm]{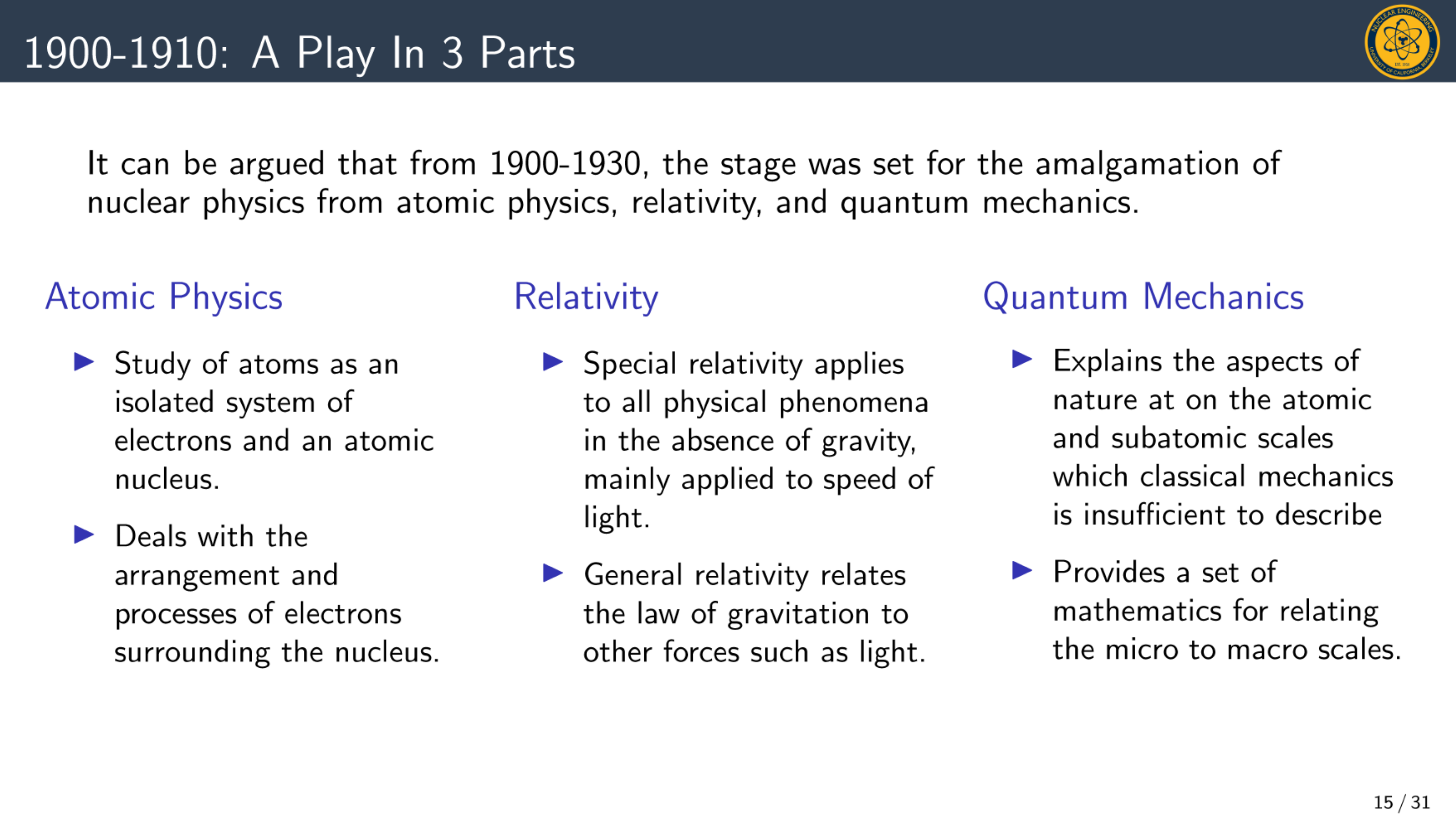}
    \caption{A Play in Two parts. Lecture W2L3.}
    \label{fig:W2L3-3parts}
\end{figure}

We then transitioned into the crux of the initial unit on \nth{20}-Century physics with the learning outcomes organized such that students would be able to:
\begin{itemize}
    \item[$\mathfrak{L}_{1}$] \textbf{Recall} the major historical milestones in early \nth{20}-Century physics and describe the experiments that led to them.
    \item[$\mathfrak{L}_{2}$] \textbf{Organize} the events on a timeline.
    \item[$\mathfrak{L}_{3}$] \textbf{Draw} connections between the developments in atomic physics, relativity, and quantum mechanics and explain how their roots in nuclear physics.
\end{itemize}
The beginning of this unit focused on setting the stage of physics (1890-1899) and understanding the ``play in 3 parts'' (1900-1910) between quantum mechanics, atomic physics, and relativity (1910-1930) as shown in Figure \ref{fig:W2L3-3parts}. 
Lectures dealing with the early history from 1900-1910 began with a discussion of the development of the mathematical landscape and key players that provided the historical context for later revelations in nuclear, atomic, quantum, and relativistic theories. Among these were Georg Cantors (1845-1918) standardization of mathematics through set theory\cite{Cantor1879UeberPunktmannichfaltigkeiten}, David Hilbert's (1862-1943) distillation of what would later be christened ``Hilbert Spaces'' by John von Neumann (1903-1957)\cite{vNeumann1930AllgemeineFunktionaloperatoren} and would prove critical in downstream developments in quantum mechanics\cite{Hilbert1928UberQuantenmechanik}, especially as it relates to Werner Heisenberg's (1901-1976) matrix methods\cite{Peres1994QuantumMethods}. 

\begin{figure}[h!]
    \centering
    \includegraphics[width=8.6cm]{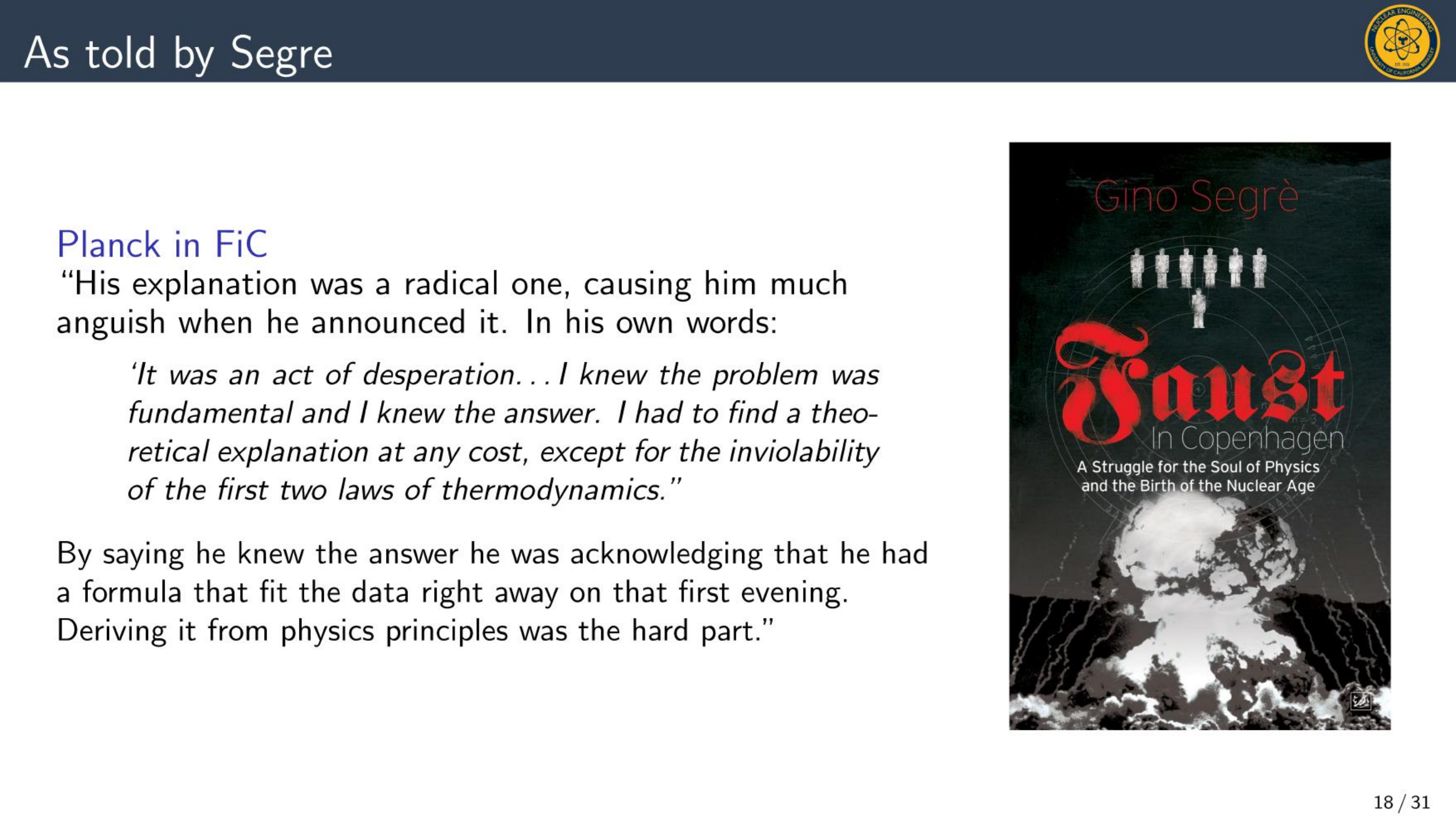}
    \caption{Planck in \textit{Faust in Copenhagen}. Lecture W2L3.}
    \label{fig:W2L3-Planck}
\end{figure}

\begin{figure}[h!]
    \centering
    \includegraphics[width=8.6cm]{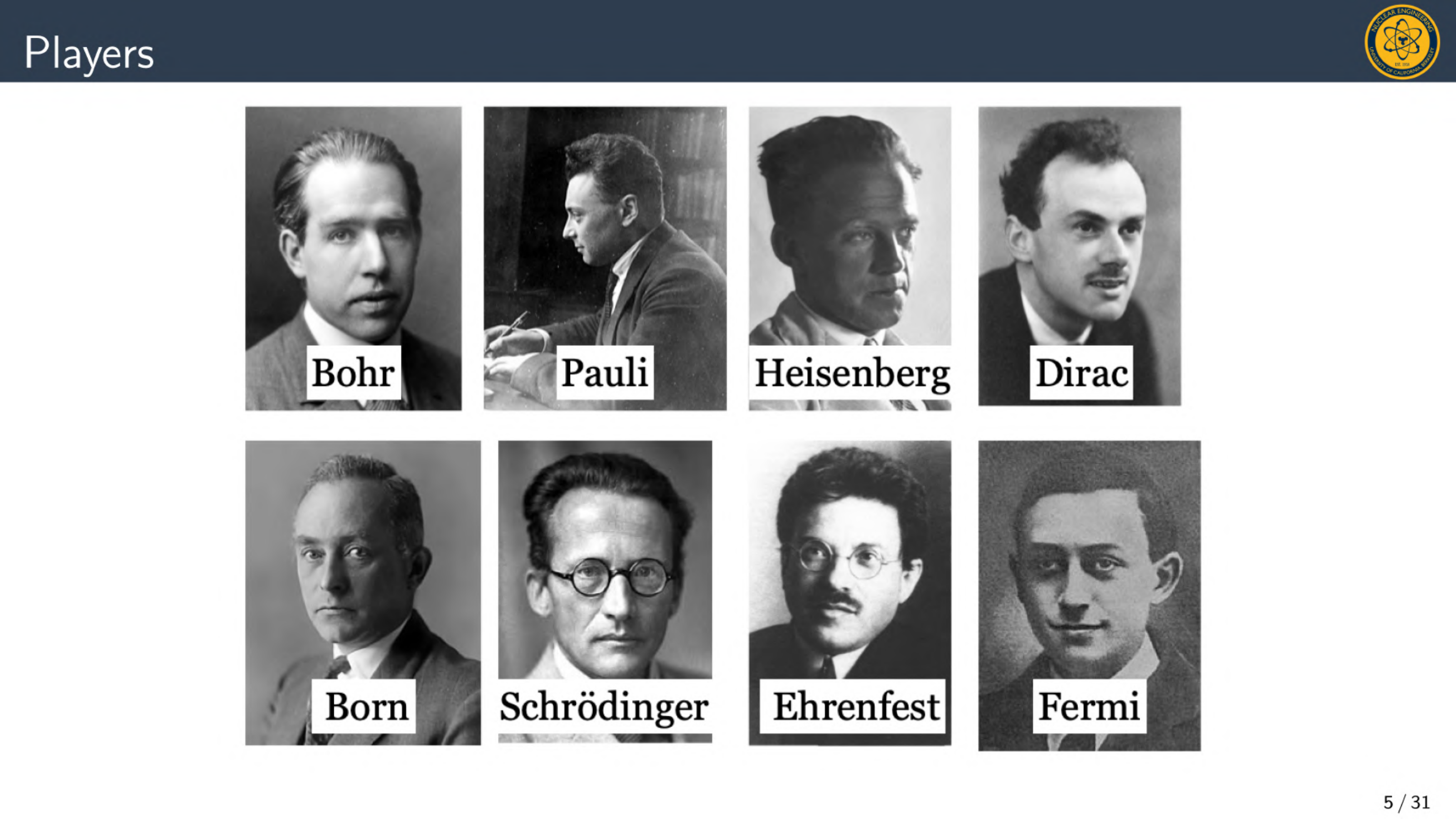}
    \caption{Important Players in \nth{20}-Century Physics. Lecture W2L4.}
    \label{fig:W2L3-players}
\end{figure}

With this requisite appreciation of the mathematical underpinnings taken care of, we introduced quantum mechanics through the works of Max Karl Planck (1858-1947) and his interest in addressing a 1859 question from Kirchoff, ``how does the intensity of the
electromagnetic radiation emitted by a black-body depend on the frequency of the radiation and the temperature of the body?'' Using Planck as the primary figure allowed us to begin discussions of the first primary reading in the form of Segre's \textit{Faust in Copenhagen}\cite{Segre2007FaustPhysics} (Figure \ref{fig:W2L3-Planck}) and the framing of the importance of historical figures at scientific gatherings such as the Solvey conferences (Figure \ref{fig:W2L3-players}).

Our goal in these early NE290 lectures was to tell the nuclear history through tales of the physicists who played a central role in its development. In order to augment this history, we used the tale of Goethe's \textit{Faust}\cite{Goethe2021Faust} to portray the dual nature of nuclear technologies. The works by Goethe and Segre set the stage as literary basis and historical yarn respectively for our introduction of Max Delbrück (1906-1981) 1932 \textit{The Blegdamsvej Faust} -- as translated by George Gamow (1904-1968)\cite{Gamow1966ThirtyPhysics}. Reserving class-time for students to act out (over zoom) selected passages from the parody, we aimed to provide an immersive environment where \nth{21}-centry nuclear engineering students in California could don the mantle of \nth{20}-century physicists on a make-shift Swedish stage who themselves were playing roles of \nth{16}-century personas. We then built on this physics-as-theater concept to relate previously introduced names from earlier lectures to the outstanding full cast of prominent \nth{20}-century scientists. Here, the fable allowed us to trace the obsession of the neutron by Pauli and Bohr. Fittingly, \textit{The Blegdamsvej Faust} ends with the neutron's discovery by James Chadwick (1891-1974) cast as Wagner -- heralding the transition from scientific discovery to wartime use, and we outlined this subtle point with a discussion of Lise Meitner's (1878-1968) attendance of the performance, but lack of participation amongst the cast. The use of \textit{The Blegdamsvej Faust} was developed throughout NE290 both as an emphasis of the complex bargains inherent to nuclear physicists and as a lens for exploring nuclear history in literature.

\subsection{Physics of WWII}

\begin{figure}[b!]
    \centering
    \includegraphics[width=8.6cm]{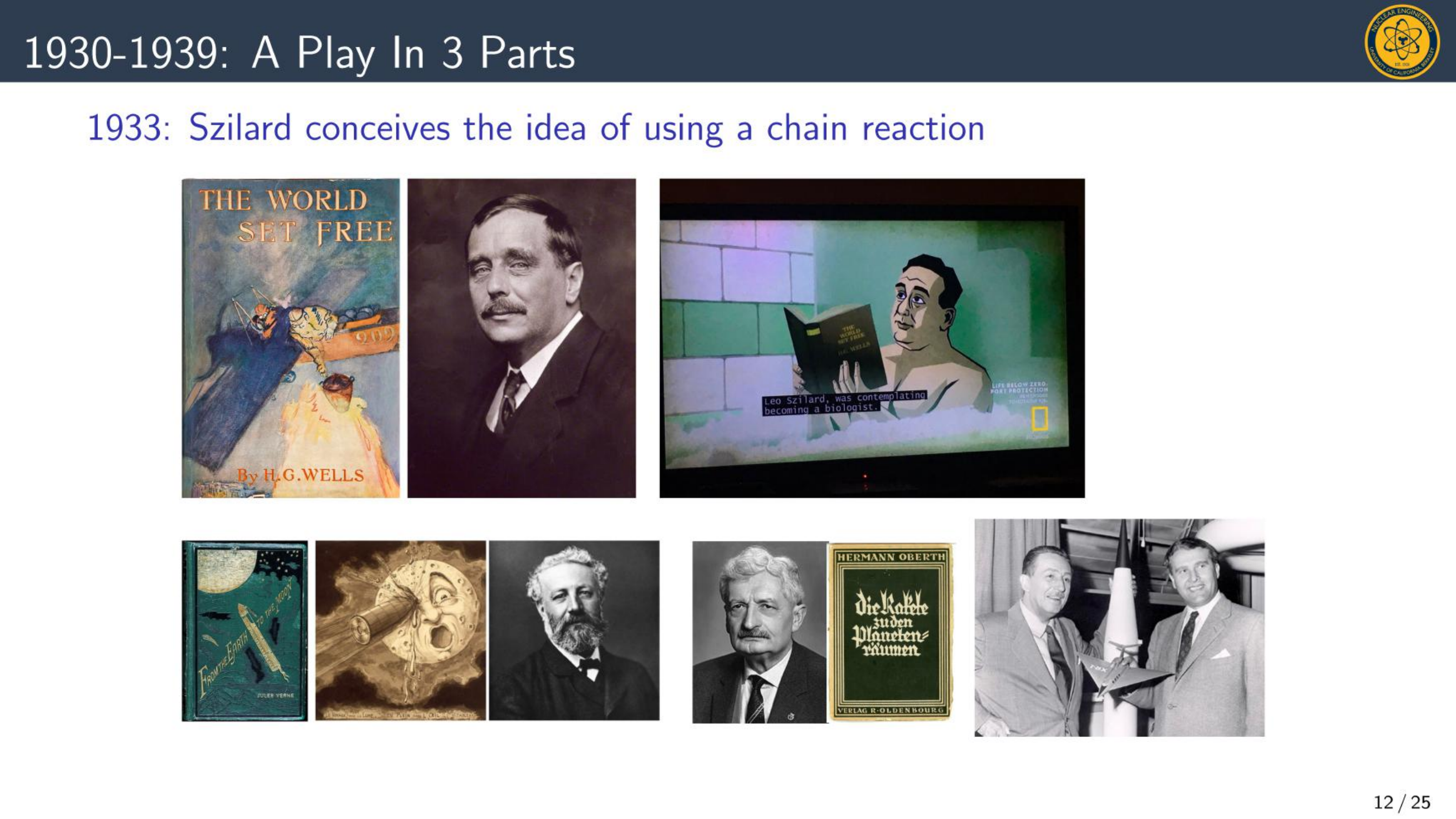}
    \caption{Slide detailing Leo Szilard's conceiving of a nuclear chain reaction following insight from H.G. Well's \textit{The World Set Free} Similar scientific insights from adjacent disciplines also shown. Lecture W4L7.}
    \label{fig:W4L7-Szilard}
\end{figure}

Following the establishment of critical aspects of \nth{20}-century physics, we continued NE290 with a unit that traced the foundations, establishment, and exploitation of nuclear energy across World War II (WWII). With similar learning outcomes ($\mathfrak{L}_{1,2}$) to the previous unit, we aimed for students
\begin{itemize}
    \item[$\mathfrak{L}_{3}$] \textbf{Draw} connections between the developments in \nth{20}-century physics and latter Manhattan project.
\end{itemize}
Like we did with our 1927 introduction of cyclotron technologies\cite{Telegdi1998SzilardMore}, we began with Szilard's 1933 conceiving of the nuclear chain reduction as an insight garnered from reading H.G. Well's \textit{The World Set Free}\cite{Wells1914TheFree} in a bathtub\cite{Ottaviani2001Fallout:Bomb} (Figure \ref{fig:W4L7-Szilard}). From here, our initial lecture outlined the proceeding events that vindicated Szilard and his idea starting with the Ida Noddack's (1896-1978) 1934 proposal for fission and observation of Neptunium\cite{Noddack1934Uber93}, Otto Hahn's (1879-1968) 1938 observation\cite{Hahn1939UberErdalkalimetalle}, and Meitner's 1938 synthesis\cite{Meitner1939DisintegrationReaction} -- culminating in Fermi's 1938 Nobel Prize and his collaboration with Szilard in realizing \textit{2n} neutron production\cite{Anderson1939NeutronUranium}. Framing these achievements in the context of a 1930s world on the brink of war, our initial lecture ended by recounting the 1939 story of Teller and Szilard lost in upstate New York -- searching for Einstein and his signature on the letter to the U.S. President that would mark the start of the Manhattan Project. 

\begin{figure}[h!]
    \centering
    \includegraphics[width=8.6cm]{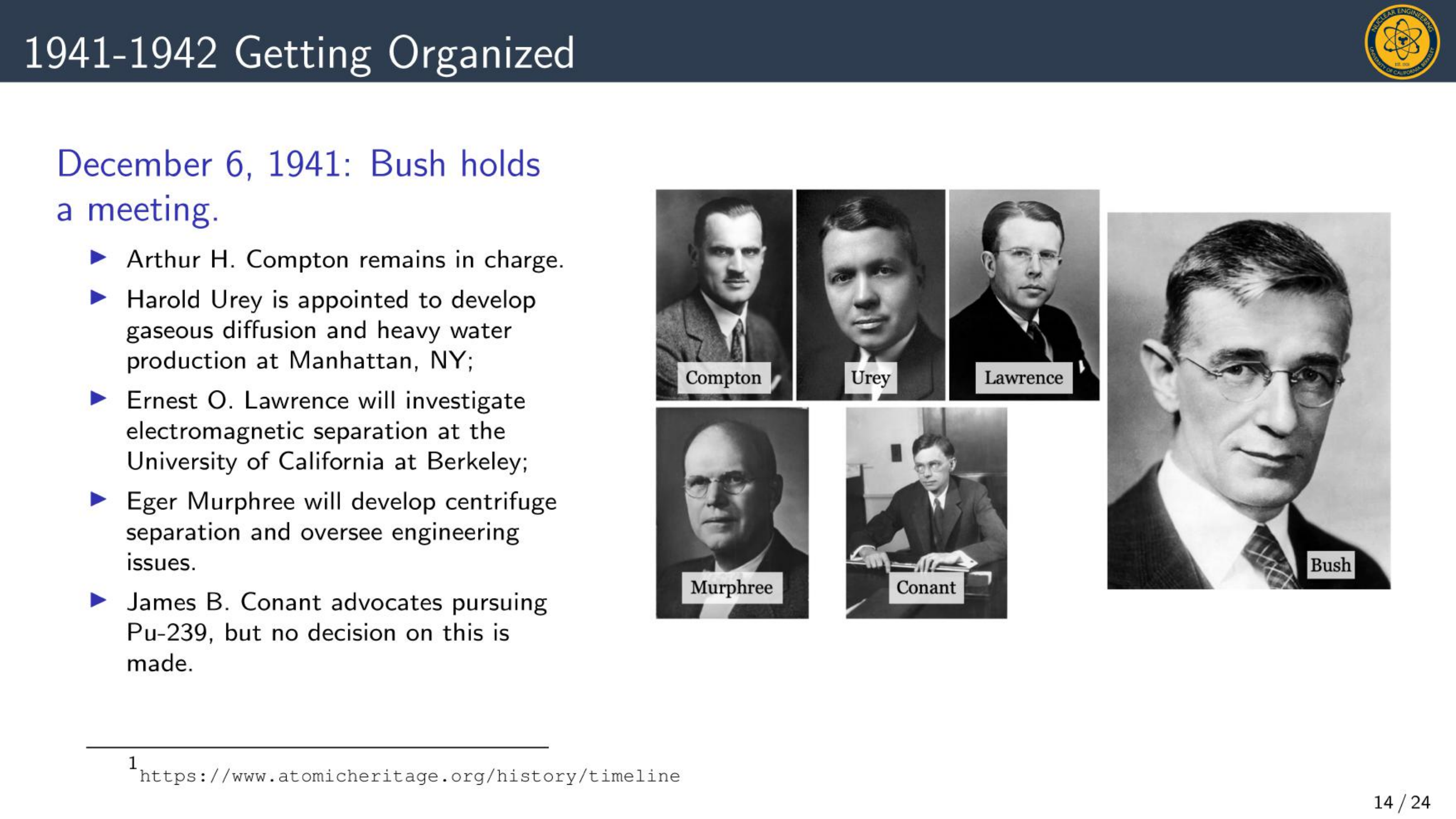}
    \caption{Slide detailing organizational leadership and structure of early MED. Lecture W4L8.}
    \label{fig:W4L8-committee}
\end{figure}

Throughout the remaining lectures in this unit (W4L8-W7L13), we primarily focused on guiding NE290 students through the timeline of events from 1939 to 1945 -- outlined in large part by -- and with our considerable appreciation to -- the \textit{Atomic Heritage Foundation}\footnote{\url{https://www.atomicheritage.org/}}. In W4L8, we first outlined the 1939-1941 investigations of nuclear weapons through the cross-talk between the U.S. Advisory Committee on Uranium, U.K. consideration of $^{235}$U fast fission by Otto Frisch and Rudolf Peirls, and the interplay between the MAUD and Briggs Committees. We then outlined the 1941-1942 Allies' efforts to organize with emphasis on the establishment of Office of Scientific Research (OSRD) and later formation of the S-1 and its scientific pillars (Figure \ref{fig:W4L8-committee}). Splitting this lecture, we then provided a technical guide on fissile  $^{235}$U purification -- as outlined originally by the S-1 committee -- via liquid thermal diffusion, gaseous thermal diffusion, electromagnetic separation using the Calutron. 

\begin{figure}[h!]
    \centering
    \includegraphics[width=8.6cm]{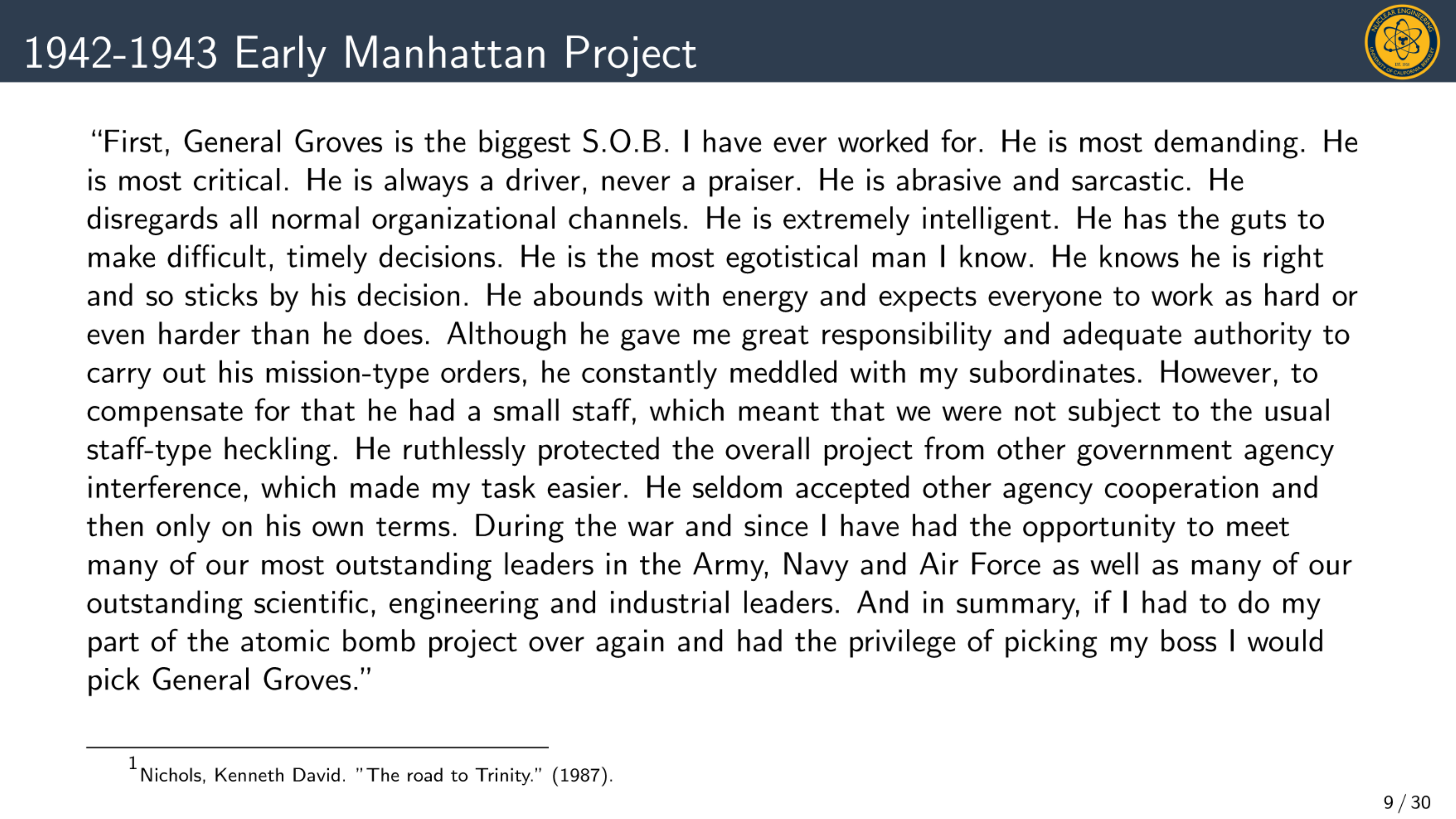}
    \caption{Slide recounting the character of General Groves by Kenneth Nichols\cite{Nichols1987TheTrinity}. Lecture W5L9.}
    \label{fig:W4L9-groves}
\end{figure}

In W5L9, we continued our lessons on the early Manhattan Project of 1942-1943 beginning with a portrait of the military overseer of the Manhattan Engineer District (MED), General Leslie Richard Groves Jr., drawn from a number of historical accounts\cite{Nichols1987TheTrinity} (Figure \ref{fig:W4L9-groves}). The character of Groves was juxtaposed in discussion when reintroducing MED scientific director J. Robert Oppenheimer to the class -- building on the characterization of Oppenheimer as a young idiosyncratic graduate student from the pages of \textit{Faust in Copenhagen}. Given the gravity of their achievements, we dedicated the entirety of W10L19 to Oppenheimer and we made use of a variety of educational mediums beyond biographies\cite{Bird2005AmericanOppenheimer,Conant2006109Alamos} such as graphic novels\cite{Ottaviani2001Fallout:Bomb}, plays\cite{Goodchild1983Oppenheimer:Bomb}, and even an opera\cite{Adams2008DoctorActs} to ensure a unique and complete accounting befitting America's first scientific superstar\cite{Oppenheimer1948TheApprentice}. In contrast to the historical and literary aspects, we also aimed to frame the story in conjunction with the technical physics for which we provided reading and discussion of the \textit{Los Alamos Primer}\cite{Serber1943LosPrimer}. 

In lectures W6L12 (Developing the Bomb) and W7L13 (Ending the War) we concluded the historical timelines for the unit. Our slides demonstrate our effort in portraying the complexity of outcomes from America's first use of the Atomic Bomb across a variety of factions that included the Japanese victims, the men and women of MED, and the sociopolitical operatives across the military and government hierarchy. In W6L11, guest lecturer Dr. Alex Wellerstein led the class in exploring the morality of first-use, making use of a number of articles that resolved misconceptions and ``set the historical record straight''\cite{Wellerstein2015Nagasaki:Bomb,Wellerstein20203.Hiroshima}.

\begin{figure}[t!]
    \centering
    \includegraphics[width=8.6cm]{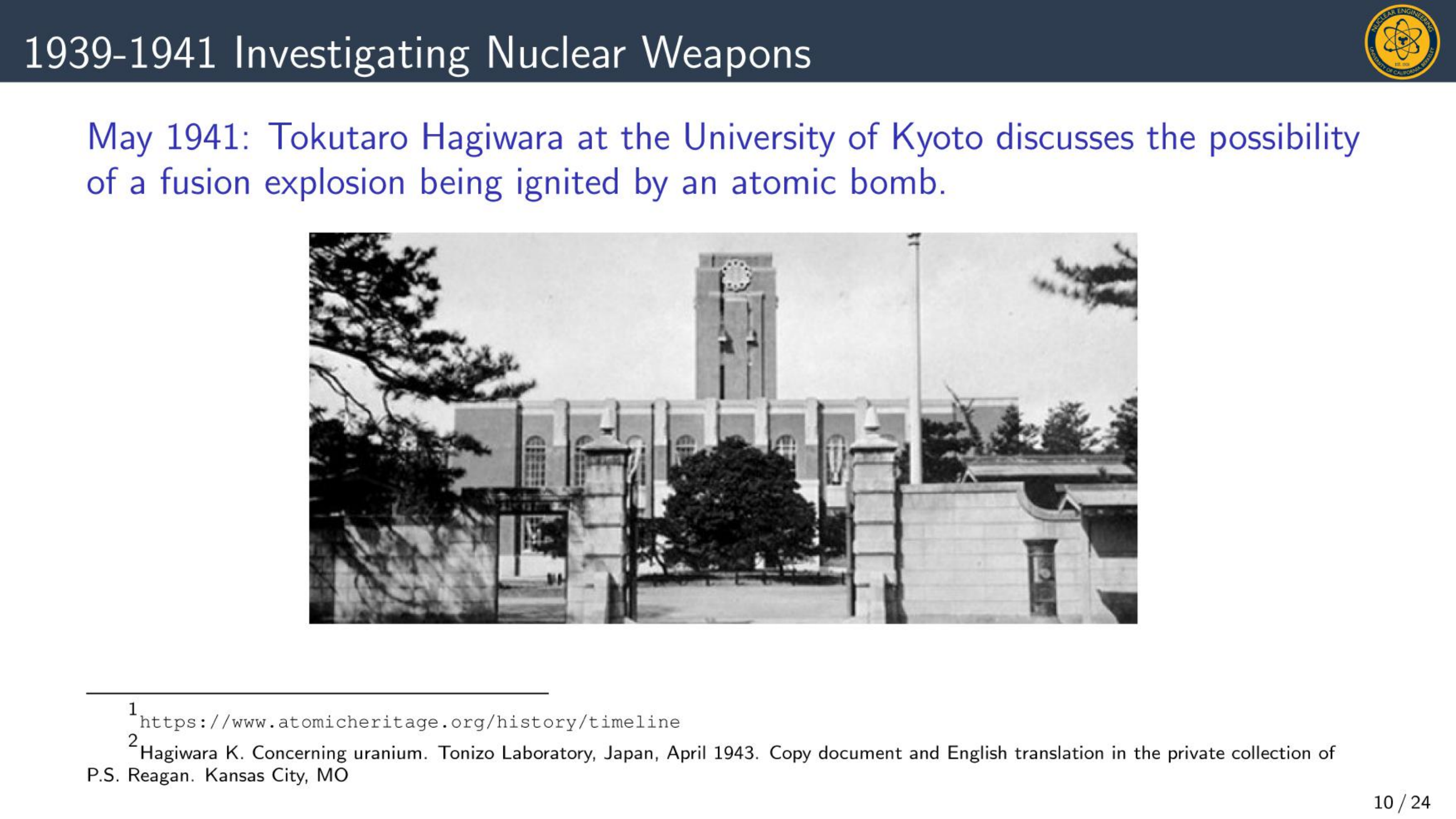}
    \caption{Slide recounting earliest conceptualization of a thermonuclear weapon in Japan. Lecture W4L8.}
    \label{fig:W4L8-hagiwara}
\end{figure}

\begin{figure}[h!]
    \centering
    \includegraphics[width=8.6cm]{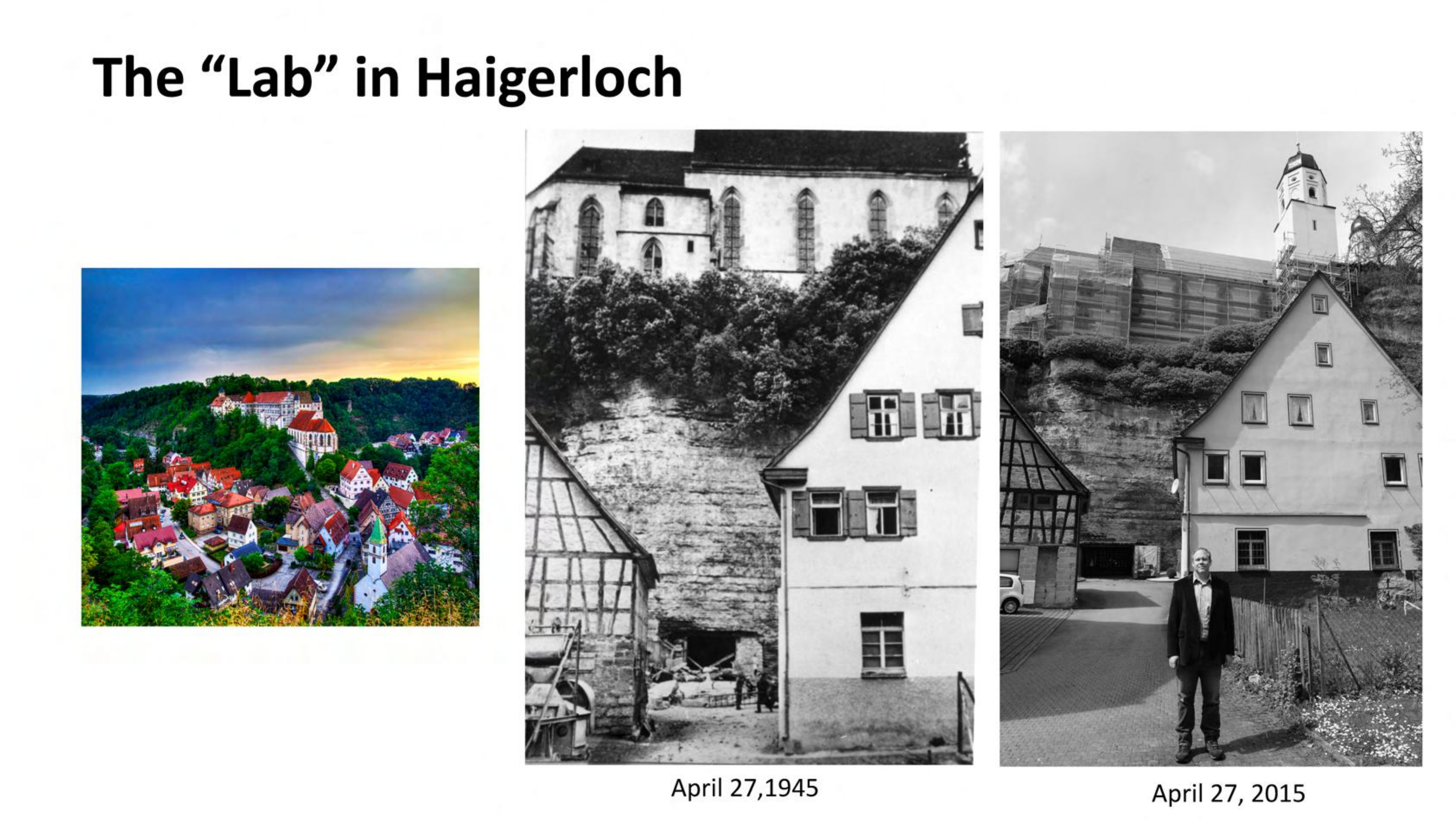}
    \includegraphics[width=8.6cm]{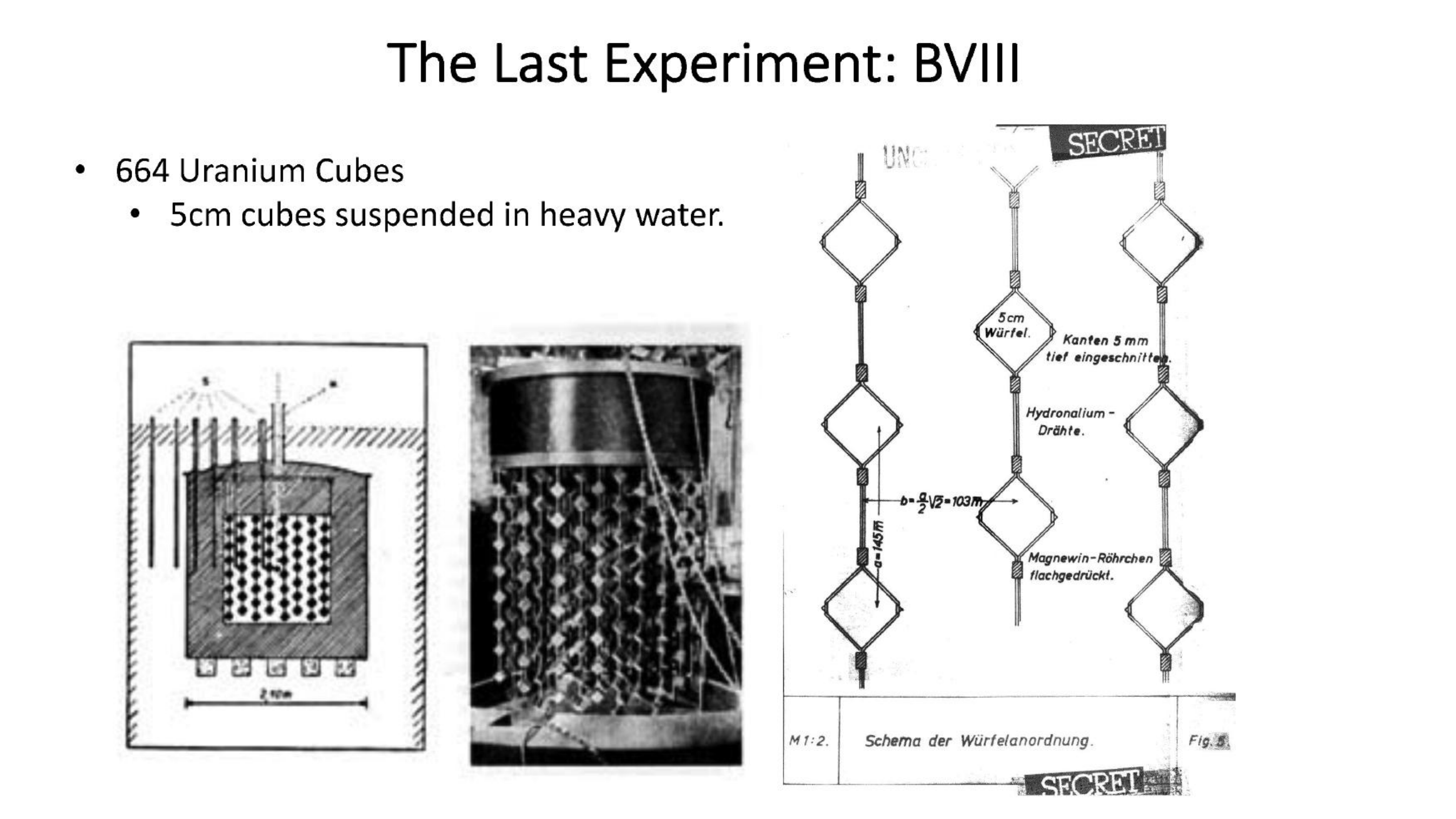}
    \caption{Slides depicting Nazi Uranprojekt laboratory led by Heisenberg in Haigerloch and experimental setup of the BVII reactor. Lecture W5L10.}
    \label{fig:W5L10-lab}
\end{figure}

Although the focus of this unit was the U.S. led MED efforts, we also endeavored to provide an accountancy for the nuclear wartime weapons programs mounted by Germany, Japan (\textit{Ni-Go}), and the Soviet Union (\textit{Sovetskiy proyekt atomnoy bomby}). The historical record of nuclear weapon development was written by scientists from all corners of the globe, and we endeavored to showcase this global effort whenever possible (Figure \ref{fig:W4L8-hagiwara}). While significant events in non-U.S. programs can be found integrated into the timeline presented in lectures, W5L10 guest lecturer Dr. Miriam Hiebert provided an in-depth analysis of the Nazi Uranprojekt (``Uranium Project'') and the U.S.-led Alsos Mission tasked with scientific intelligence gathering (Figure \ref{fig:W5L10-lab}). Here the NE290 class was reintroduced to an older Heisenberg -- who, like Oppenheimer, was scripted into \textit{The Blegdamsvej Faust}. Following this lecture, the class was provided the audiobook of Michael Frayn's play \textit{Copenhagen}\cite{Frayn2017Copenhagen} (with voice acting by Benedict Cumberbatch) based on the 1941 meeting in Copenhagen between the physicists Niels Bohr and Heisenberg -- and provides a literary window into the circumstances for and ramifications of Heisenberg's place in history. Additionally Hiebert's lecture introduced the class to technical engineering literature which applied modern nuclear phyiscs simulations to reconstructed Nazi-era reactors\cite{Grasso2009AReactor}. Heibert's discussion of the Alsos mission led into our final lecture W7L14 detailed the sharing of nuclear data (voluntary and otherwise). Here we outlined a number of cases of Soviet spy-craft occurring throughout the wartime MED, and ushering the class into the following unit on the early Cold War. 

\subsection{Early Cold War}
The early cold war section of the course focused on the US efforts to build a nuclear arsenal and consider the conditions under which it would be used. This included the formation of the AEC, the provisions of the atomic energy act, the formulation of the first nuclear strategies and the evolution of command and control. 
\begin{figure}[h!]
    \centering
    \includegraphics[width=8.6cm]{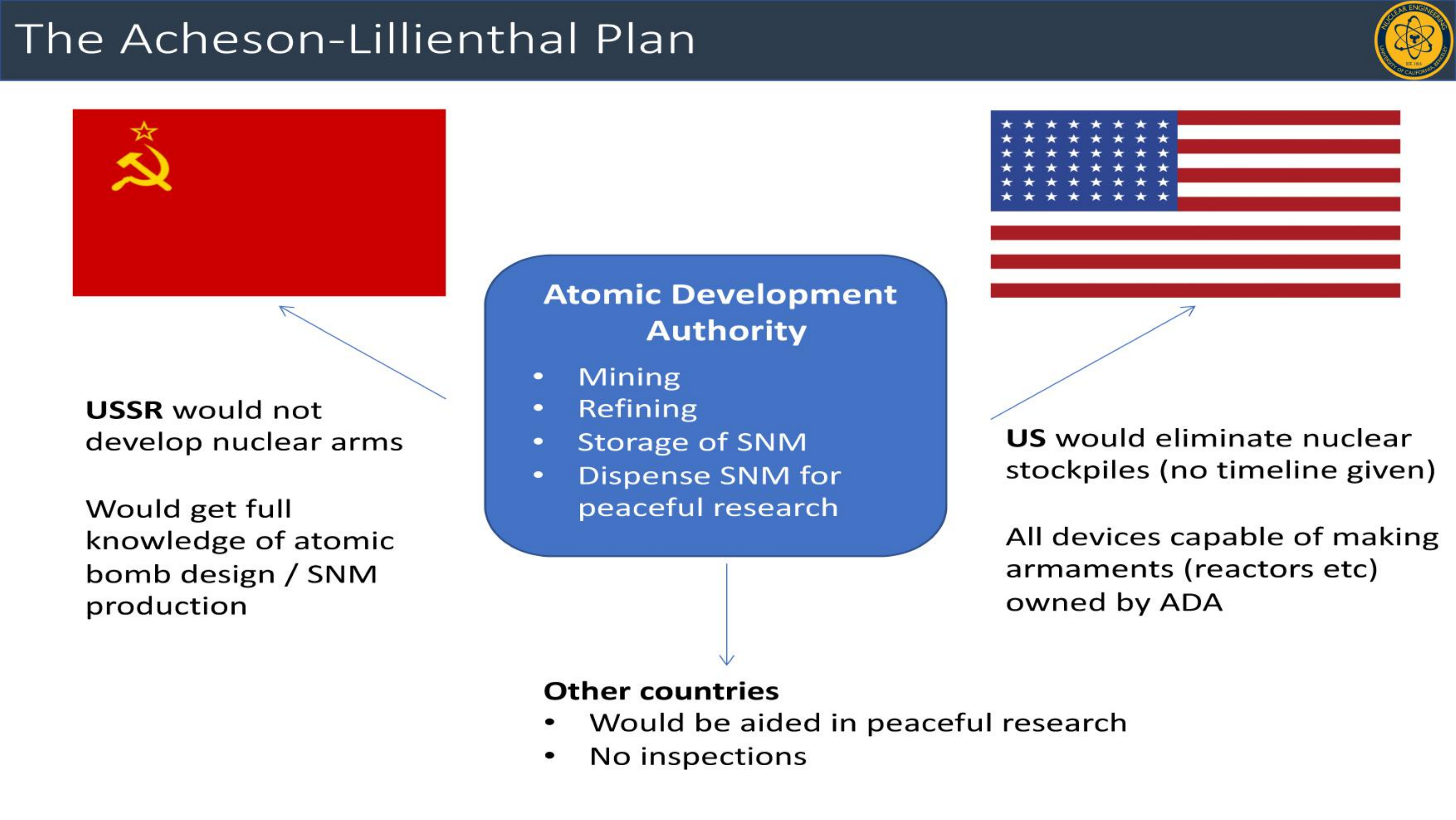}
    \includegraphics[width=8.6cm]{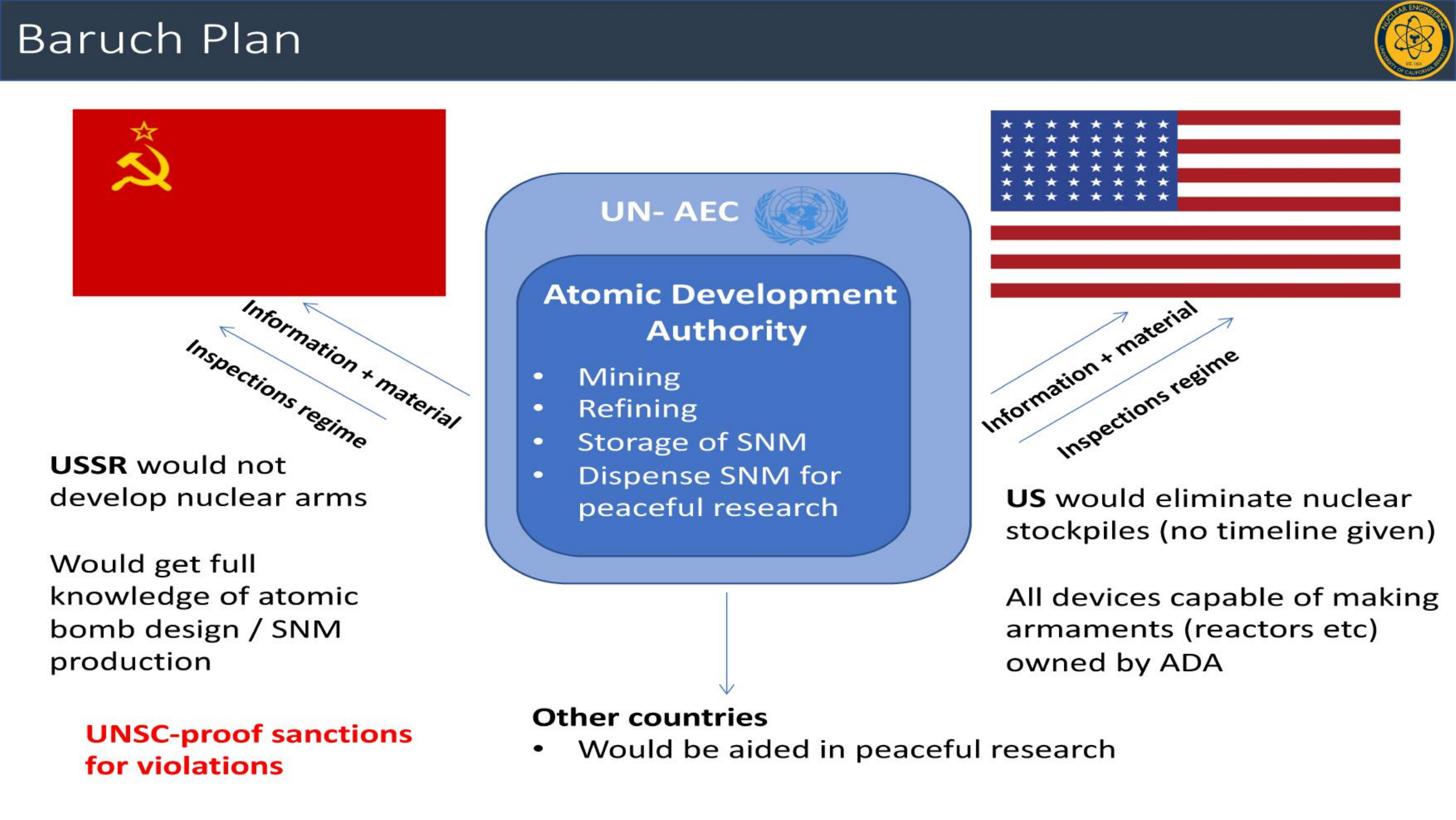}
    \caption{Slides outlining the Acheson-Lillienthal (top) and Baruch (bottom) Plans. W7L14.}
    \label{fig:W7L14-plan}
\end{figure}
\begin{figure}[h!]
    \centering
    \includegraphics[width=8.6cm]{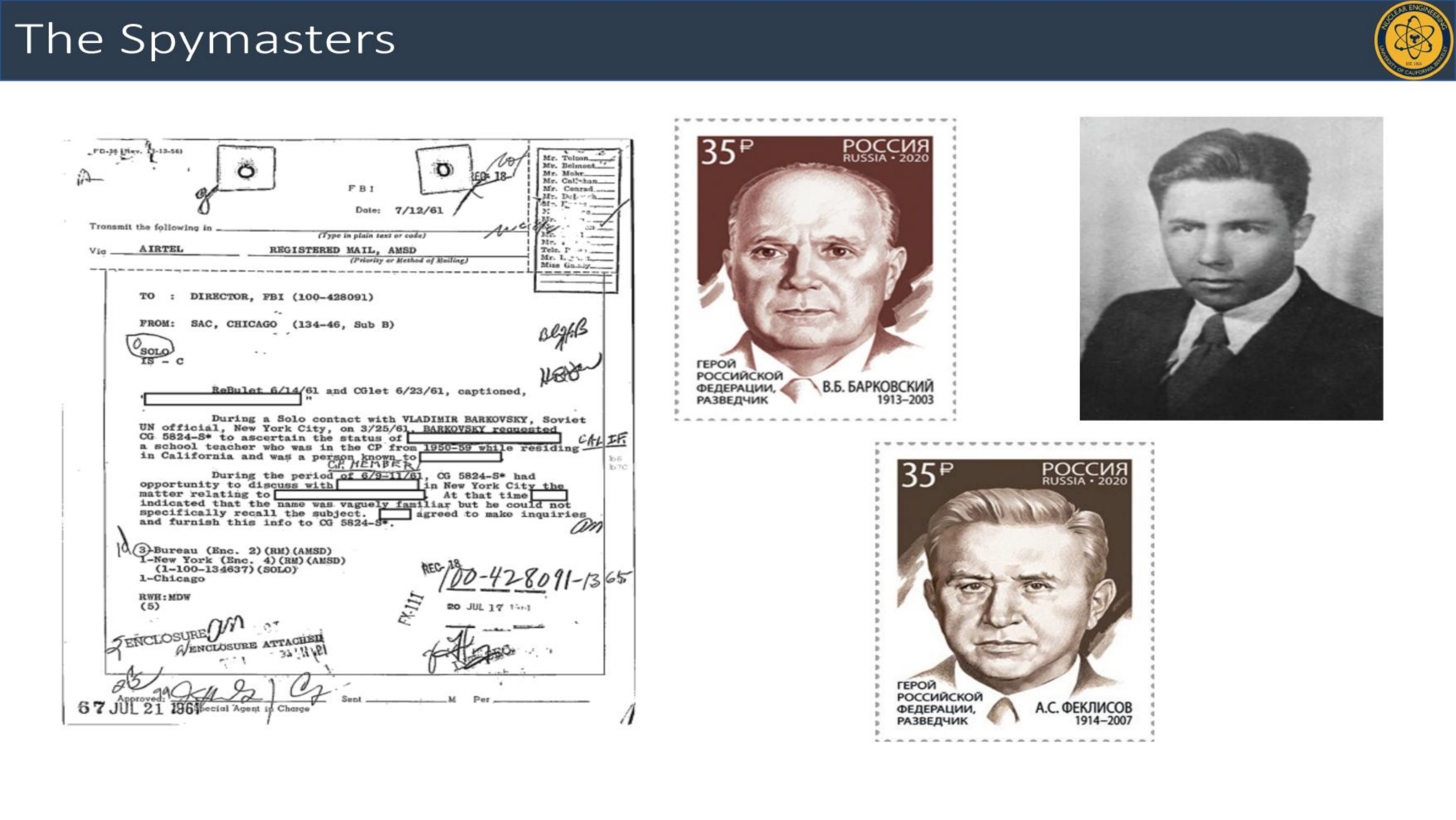}
    \caption{Slides describing Soviet spymasters. W7L14.}
    \label{fig:W7L14-spy}
\end{figure}
Topics included the Berlin crisis, massive retaliation, the Castle Bravo test and the beginning of global anti-nuclear activism. On the Soviet side, topics included the Soviet nuclear program, starting from Flyorov's letter to Stalin, to the test of RDS-1 in 1949 and RDS-37 in 1955. This series of lectures were augmented with readings from declassified documents that aided students in understanding the gulf between perceptions and the military reality in this era. 

This unit provided an opportunity to coordinate classroom discussions on topics of public nuclear policy such as the Acheson-Lillienthal Plan as compared to the Baruch Plan (Figure \ref{fig:W7L14-plan}). Such discussions allowed the students follow the historical record through the lens of public response to the nuclear energy. Additionally, we contrasted the history of public nuclear policy with lectures on the private aspects of early Cold War nuclear spycraft (Figure \ref{fig:W7L14-spy}). 

\begin{figure}[t!]
    \centering
    \includegraphics[width=8.6cm]{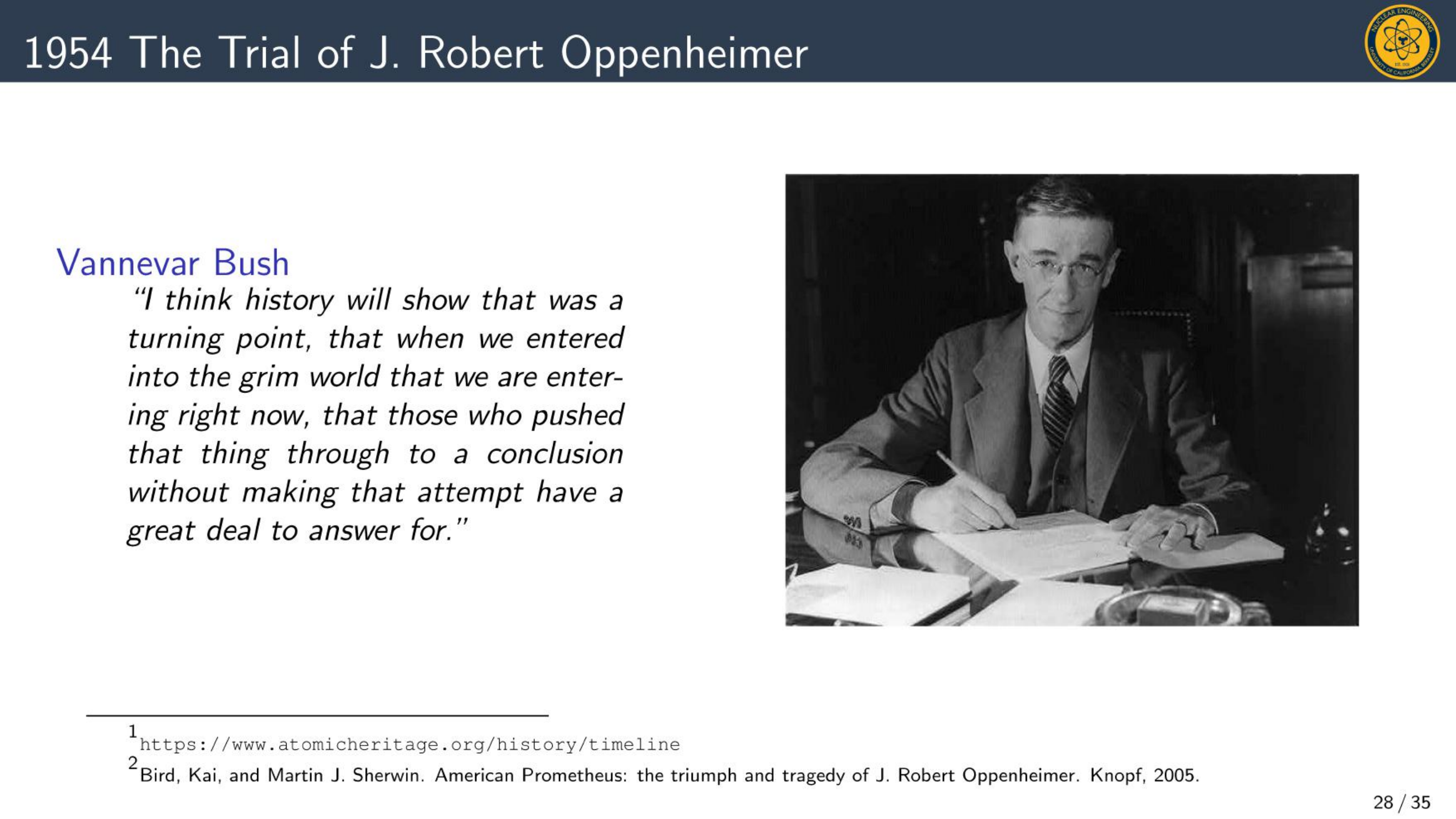}
    \includegraphics[width=8.6cm]{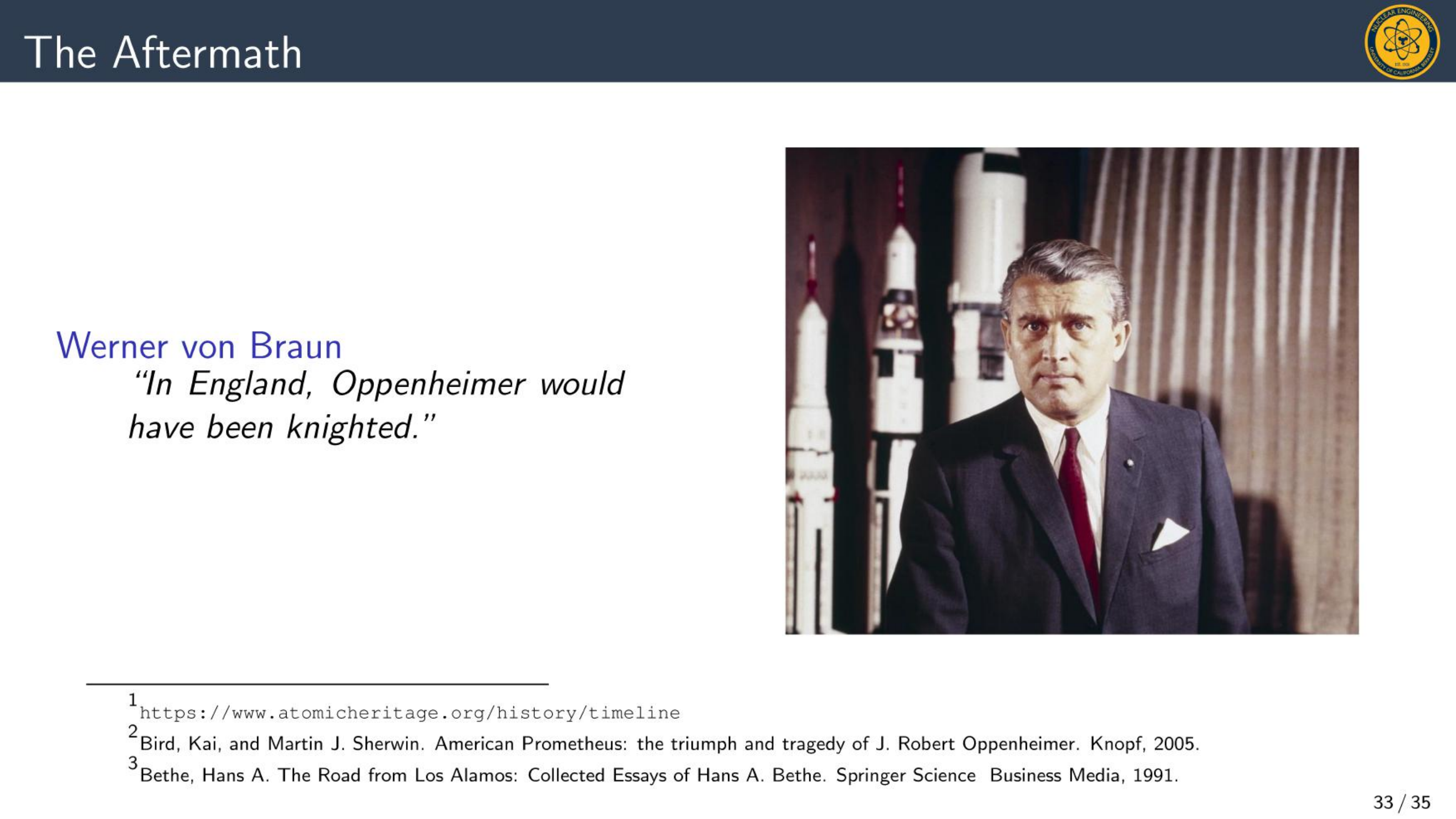}
    \includegraphics[width=8.6cm]{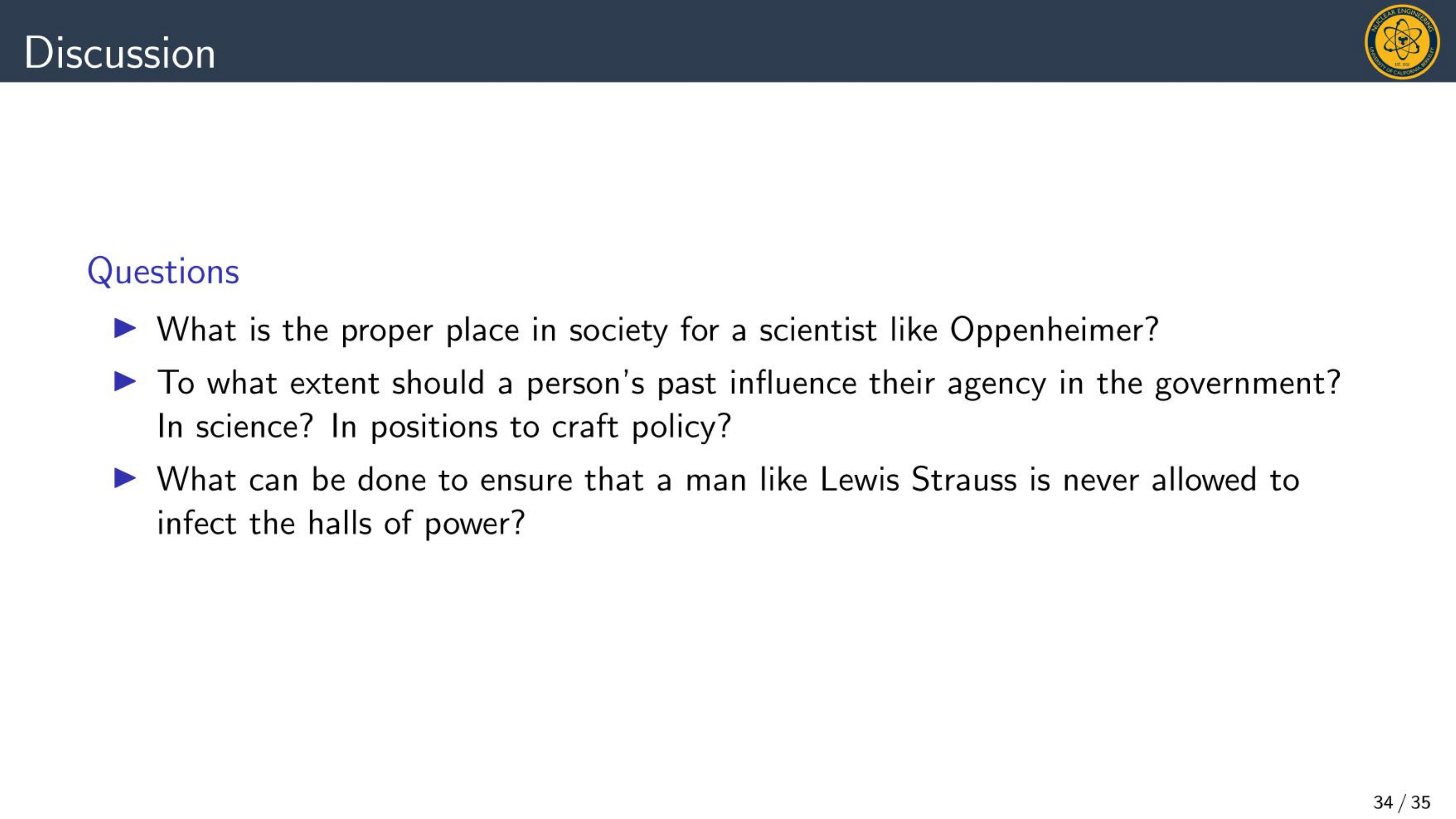}    
    \caption{Slides outlining the public and scientific response to the trial of Robert Oppenheimer. W10L19.}
    \label{fig:W10L19-opi}
\end{figure}

In addition to the technical and international foreign affairs aspects of the early Cold War, we dedicated an entire lecture to examining the 1954 trial of Robert Oppenheimer. We began this lecture with his departure from Los Alamos in 1945 and his subsequent 1947 move to Princeton -- setting the stage for Oppenheimer's rise and fall as a public figure. We provided a historical accounting of the political tensions arising from Oppenheimer's chairmanship of the Atomic Energy Commission (AEC) and his difficult relationship to Lewis Strauss -- the man who would lead the campaign to remove his security clearance. This lecture allowed us to build on Oppenheimer's role in the previous two units from young scientist to scientific statesman and expand on the narrative of Oppenheimer as a \textit{Faustian} figure through classroom discussion (Figure \ref{fig:W10L19-opi}). 

\subsection{Late Cold War and Modern Era}
This section of the course was wide-ranging, and covered everything from the "missile gap" of the early 1960s through contemporary struggles with nuclear proliferation. Guest lecturers were brought in to cover cooperative threat reduction, civil defense, nuclear smuggling, the Iran Deal, and modern efforts to grapple with the toll of the nuclear weapons complex. 
\begin{figure}[h!]
    \centering
    \includegraphics[width=8.6cm]{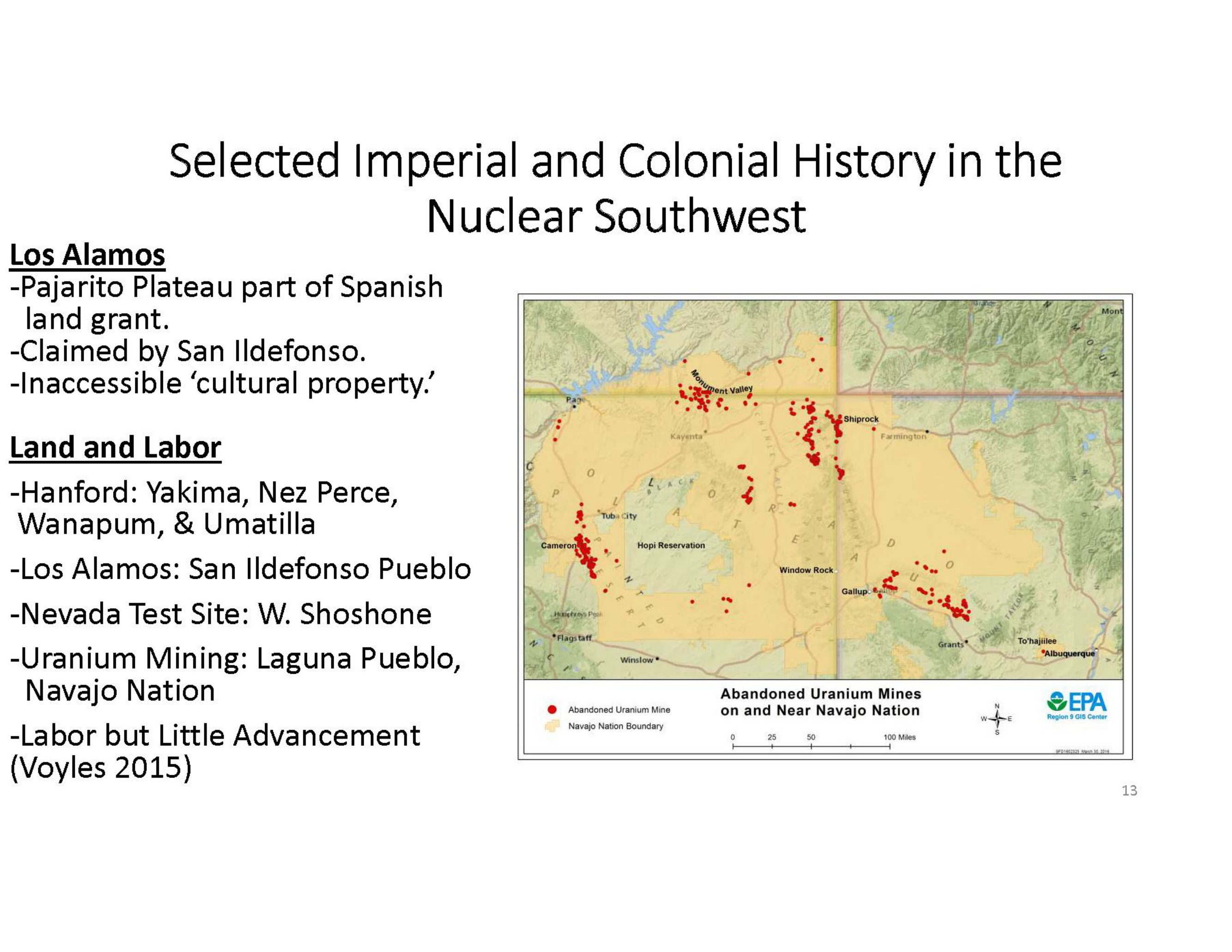}
    \caption{Slide outlining imperial and colonial history in the nuclear southwest. W11L22}
    \label{fig:W11L22}
\end{figure}
Of particular importance was the guest lecture by Marty Pfeiffer on Nuclear Colonialism which prompted students to address a number of challenging sociological and ethical considerations inherent to Nuclear Engineering at large. This course material was unique in that people involved directly in these events were lecturing, connecting the current generation of nuclear scientists to those who made the history they must live with. In comparison to earlier lectures, these dealt with topics that do not have settled interpretations. In particular, we had the opportunity to explore varying interpretations of the Iran deal, the value of deterrence in a post-Cold War world, and the modernization of the nuclear arsenal.

\begin{figure*}
\includegraphics[width=17cm]{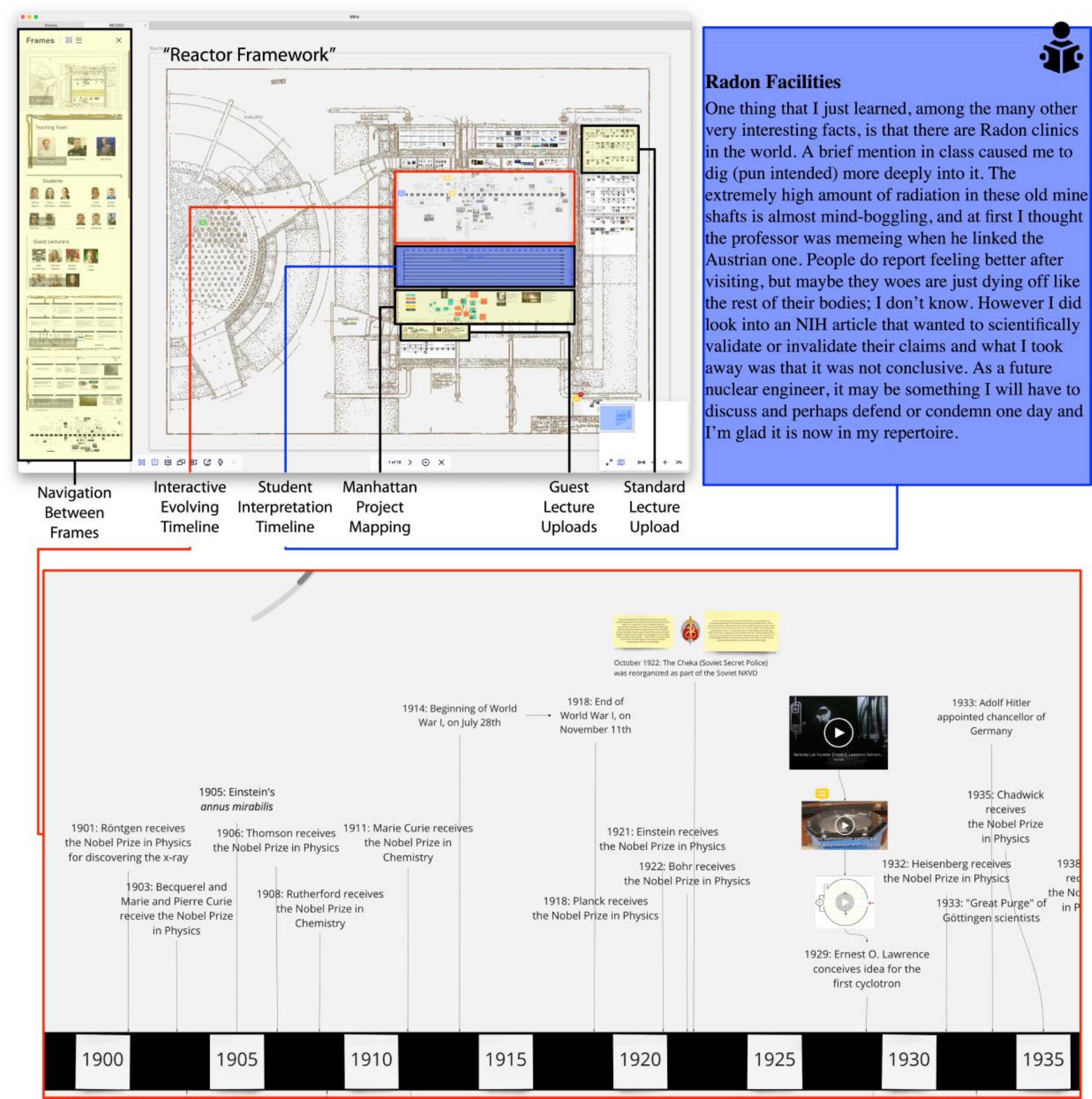}
\caption{Top left shows the complete MIRO board for NE290 and is annotated with specific features including the integrative evolving timeline, student interpretation timeline, and lecture slide integration. Top right in blue shows an example of a student response in regards to a lecture that discussed sociological issues concerning Radon contamination. Bottom shows an expanded timeline composed of milestones added by students throughout the course of NE290.} 
\label{fig:miro} 
\end{figure*}

\section{Remote Learning Environment}

The initial Spring 2021 offering of NE290 occurred during the Pandemic and thus the course was exclusively taught remotely. In order to facilitate an effective learning environment, we tailored NE290 with a number of modern tools including SLACK for communication, bCourses as the primary course file system and location for students to submit assignments and collect grades, and MIRO for interactive class collaboration on a timeline of historical events. The use of MIRO shown in Figure \ref{fig:miro}. In conjunction with the timeline of historical events, students were provided a ``response'' or ``meta'' timeline across which they added their reflection assignments. Our goal was to explore the connection between these two timelines.



\section{Term Paper}
\begin{figure}[h!]
    \centering
    \includegraphics[width=8.6cm]{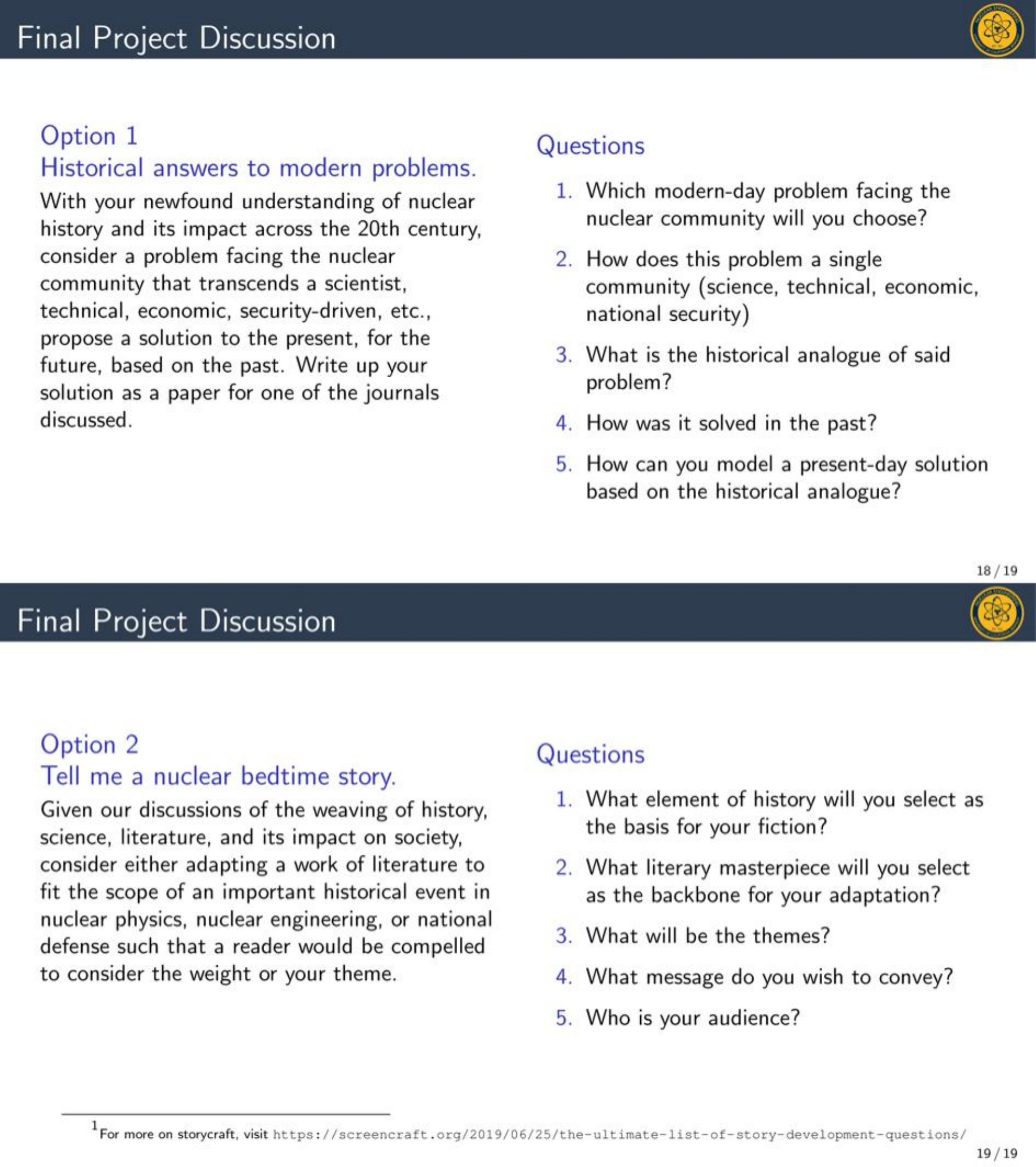}
    \caption{Final project assignment options and corresponding initial questions from Lecture W7L13.}
    \label{fig:finalassignment}
\end{figure}

In accordance with the ABET student outcomes ($\mathfrak{O}$) and the target NE290 learning goals ($\mathfrak{B}$), students were assigned a final project based on selection from two options as shown in Figure \ref{fig:finalassignment}.
In Option 1 (``Historical Answers to Modern Problems''), students were asked to consider a problem facing the nuclear community that transcends the bounds of scientific, technical, economic, or security communities, and with their newfound understanding of nuclear history and its impact across the \nth{20}-century, propose a solution to the present, for the future, based on the past. In Option 2 (``Nuclear Bedtime Story''), students were asked to consider either adapting a work of literature to fit the scope of an important historical event in nuclear physics, nuclear engineering, or national defense such that a reader would be compelled to consider the weight their chosen theme. 

In a fortunate happenstance given the differences in student preference, the class divided itself in approximately equal measure between the two options. A midterm assignment was given for each group to provide answers to the initial questions shown in Figure \ref{fig:finalassignment} which were later used for an in-class discussion and we endeavored to foster cross-talk between the two groups in the form of peer review. 

The final deliverable from the Option 1 prompt was a technosociological paper entitled ``A Nonproliferation Retrospective: How Past Successes and Failures of the Nonproliferation Regime can Inform Future Actions'' with the following synopsis:\\

\begin{myindent}
{\small
A modern-day problem the nuclear community continuously faces is how to successfully prevent proliferation. This term describes both instances of preventing nuclear states with interests in developing nuclear weapons from actively pursuing them, as well as the more difficult task of halting active development and the elimination of stockpiles. Whether through peaceful negotiations, credible threats, or use of force, a variety of strategies have been attempted throughout the course of the nuclear age with varying degrees of success. We will be undertaking the task of analyzing these precedents to piece together an understanding relating the contextual factors, historical timing, international relationships, and chosen nonproliferation approaches to the resultant response ranging from successful peaceful disarmament in some cases, to hostility and increased risk of nuclear war in others.}\\
\end{myindent}

\begin{figure}[h!]
\includegraphics[width=8.6cm]{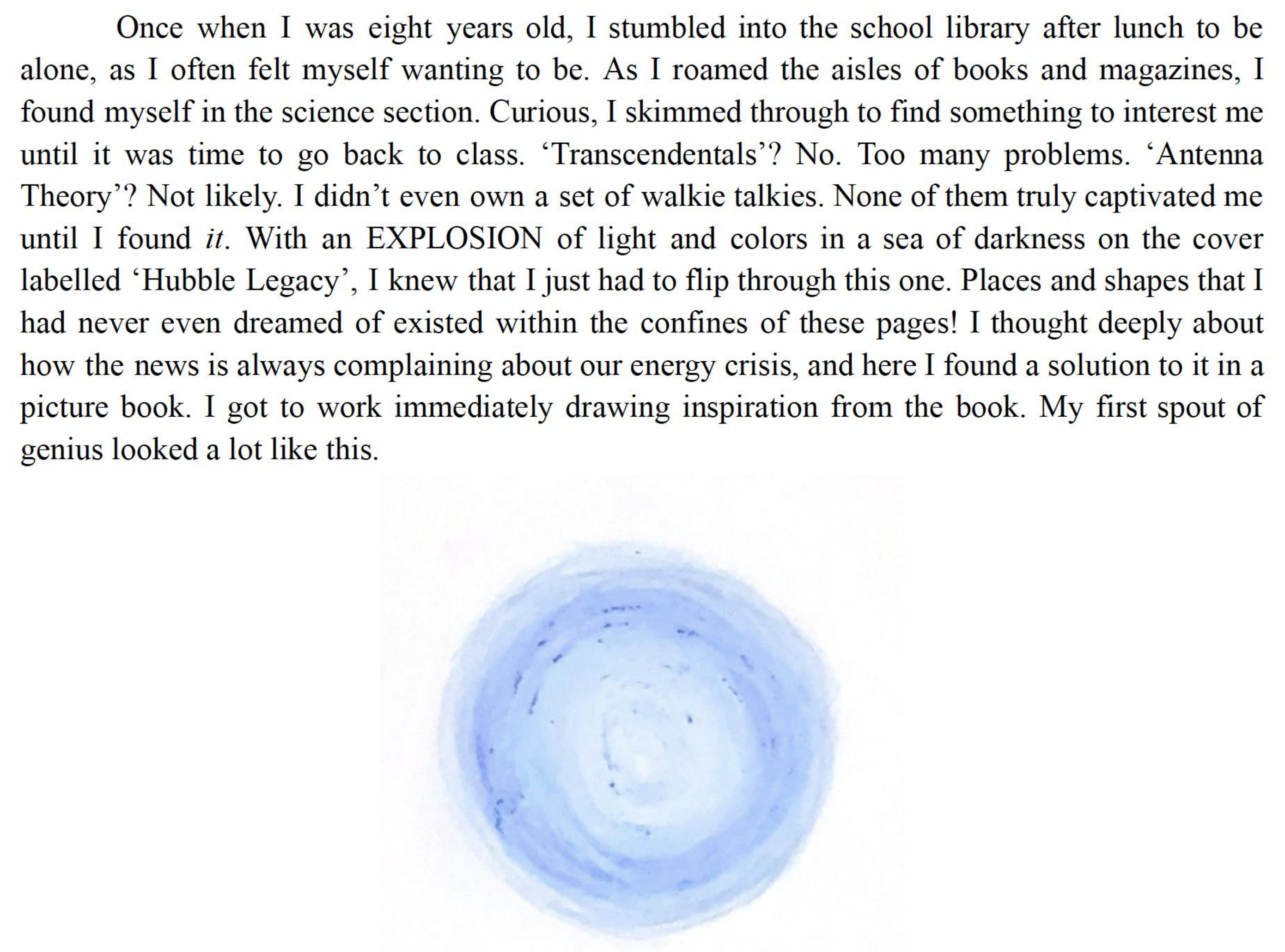}
\caption{Section of Final Project Option 2.} 
\label{fig:tls1} 
\end{figure}

\begin{figure}[h!]
\includegraphics[width=8.6cm]{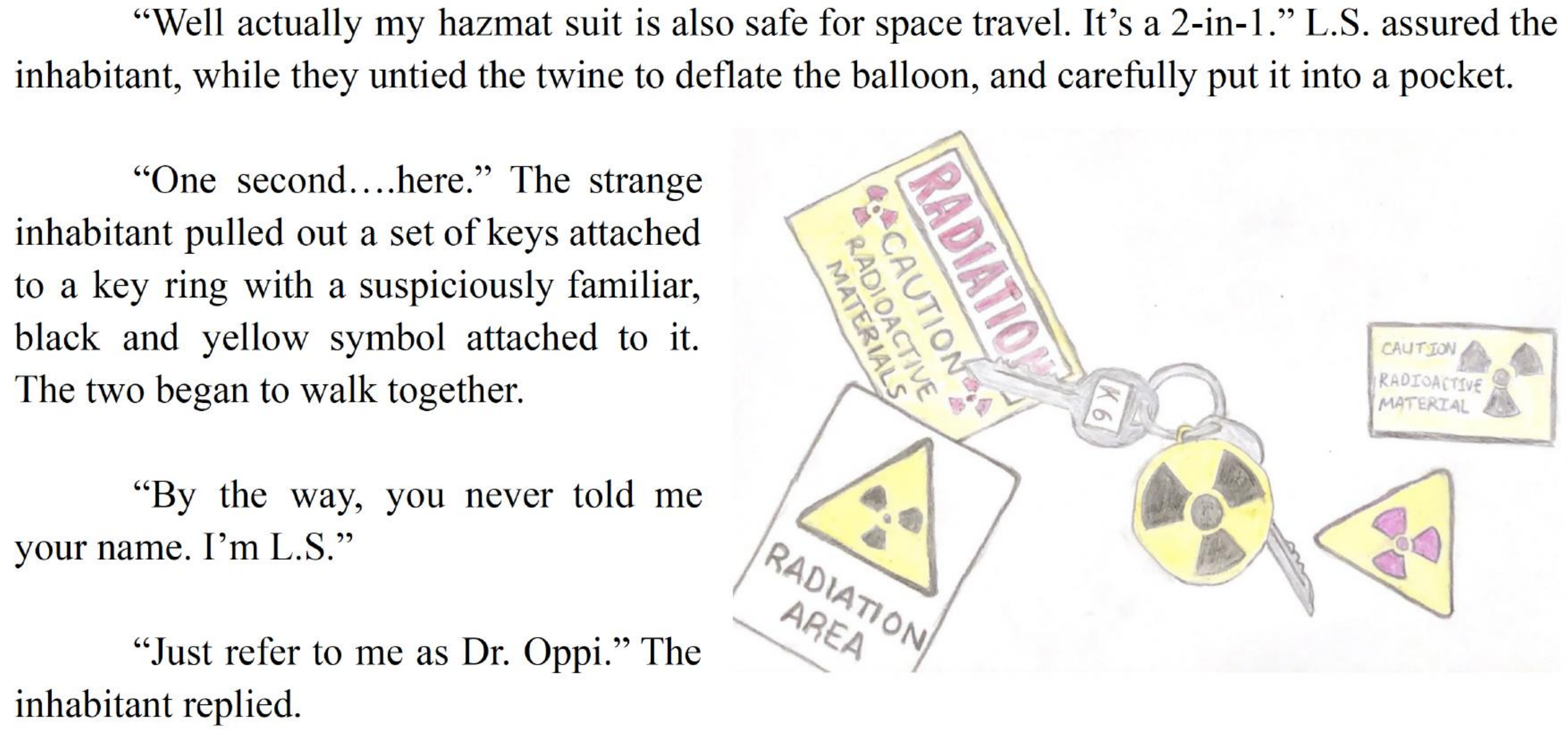}
\caption{Section of Final Project Option 2. Note the introduction of Oppenheimer as a character in the narrative.} 
\label{fig:tls2} 
\end{figure}

The final deliverable from the Option 2 prompt was an illustrated story entitled \textit{The Little Scientist} based on Antoine de Saint-Exupéry's \textit{The Little Prince}. In the original book, the narrator crashes his plane in the Sahara desert and meets the titular young boy. In the re-crafted story -- the beginning of which is shown in Figure \ref{fig:tls1} -- Stranded in the desert, the narrator tries to repair his aircraft from the crash while the little prince recounts his life story. The prince is from his own planet, and he leaves to explore the universe and visits six other planets before arriving on planet Earth. 
On each planet, the prince interacts with a different character that teaches the readers of this book different things about life and adulthood. The six planets will represent a different theme of life that coincides with a different nuclear technological application. The primary resident on each planet will represent a prominent figure for that particular nuclear technology as shown in Figure \ref{fig:tls2}.


\section{Course Outcomes}
NE290 was offered at UC Berkeley during the COVID-19 pandemic spring of 2021, and so the entirety of the course was conducted remotely via the Zoom tool. The class was composed of 11 students (2 female and 9 male). In terms of background, 10 were enrolled in nuclear engineering programs; 4 were \nth{1}-year graduate students, 3 were \nth{2}-year graduate students, 1 was a \nth{3}-year graduate student, and 1 was an undergraduate junior who was enrolled studying environmental engineering. Following the completion of the course, students were provided a survey to determine the impact of lesson in terms of its pedagogical targets. First students were asked to evaluate how NE290 aided their agency towards the nontechnical ABET goals ($\mathfrak{O}$) as shown in Table \ref{tab:abet}. 
\begin{table}[h!]
\begin{tabular}{@{}lll@{}}
\toprule
$\mathfrak{O}$ & AVG        & STD        \\ \midrule
1 & 7.67 & 2.12 \\
2 & 5.67 & 3.71 \\
3 & 8.0          & 2.96 \\ \bottomrule
\end{tabular}
\caption{Calculated student survey responses addressing ABET Outcomes $\mathfrak{O}$.}
\label{tab:abet}
\end{table}
The results show post NE290, students feel they gained agency in ($\mathfrak{O}_{1}$) an ability to recognize ethical and professional responsibilities in engineering situations and ($\mathfrak{O}_{3}$) an ability to acquire and apply new knowledge as needed, using appropriate learning strategies. However, the feedback for ($\mathfrak{O}_{2}$) suggest that the strategies taken to provide additional practice on engineering teams were insufficient -- providing an important path forward for educators in preparing to offer an NE290-like course. Students were also to evaluate their abilities relating to the learning outcomes based on Bloom's taxonomy ($\mathfrak{B}$) prior to and post completion of the NE290 course. The results 
\begin{table}[h!]
\begin{tabular}{@{}lllllll@{}}
\toprule
$\mathfrak{B}$ & \multicolumn{2}{l}{PRIOR} & \multicolumn{2}{l}{POST} & \multicolumn{2}{l}{DIFF} \\ \midrule
   & AVG         & STD         & AVG         & STD        & AVG         & STD        \\
1  & 3.33        & 1.5         & 7.56        & 1.51       & 4.22        & 1.09       \\
2  & 4.67        & 2.4         & 8.22        & 1.09       & 3.56        & 1.59       \\
3  & 4.44        & 1.81        & 8.11        & 0.78       & 3.67        & 2.12       \\
4  & 3.89        & 2.03        & 7.22        & 1.86       & 3.33        & 1.94       \\
5  & 4.56        & 2.19        & 7.89        & 0.93       & 3.33        & 2.35       \\
6  & 5.22        & 1.86        & 7.78        & 1.2        & 2.56        & 2.07       \\
7  & 5.22        & 1.39        & 8.11        & 0.78       & 2.89        & 1.54       \\ \bottomrule
\end{tabular}
\caption{Calculated student survey responses addressing targeted learning outcomes $\mathfrak{B}$ based on Bloom's taxonomy.}
\label{tab:lo}
\end{table}
The results shown in Table \ref{tab:lo} show an average agency of 4.47/10 across all $\mathfrak{B}$ prior to NE290 with an average increase of $\sim$3.37 to 7.84/10 following the course. Given the novelty in providing fully-remote education, we also questioned students on the effectiveness (on a similar scale of 1-10) of using the the tools MIRO for collaborative whiteboarding and SLACK for class-wide communication and the results are shown in Table \ref{tab:tools}.
\begin{table}[h!]
\begin{tabular}{@{}lll@{}}
\toprule
Tool  & AVG  & STD  \\ \midrule
MIRO & 5.78 & 3.49 \\
SLACK & 7.33 & 3.24 \\ \bottomrule
\end{tabular}
\caption{Remote Learning Tool Survey}
\label{tab:tools}
\end{table}
When asked to select only a single tool to keep if NE290 was taught in-person, 44\% opted for recorded lectures (enabled by Zoom), 33\% selected SLACK, and only 22\% opted for keeping MIRO . These results suggest that our use of MIRO will require additional effort in integrating the software during proceeding semesters. In reviewing the free-form feedback from students, we recieved a comment noting, ``I thought that MIRO was not helpful at all. It was just too clunky, unpleasant to look at. There's too much information to pack into a nice looking timeline. However, I did like the idea of having all the class content laid out like MIRO did. The execution was off.'' However, in reviewing the outcome survey, we are pleased the report that when asked to evaluate their understanding of Nuclear History prior to and post completion of the course, the students indicated an average jump of $\sim$3.67 from 4 to 7.67. Such feedback, while unofficial, suggests that the efforts to offer the opportunity of exploring nuclear science and engineering from a historical and literary perspective was valuable from an educational perspective. 

\section*{Resource Availability}
The complete set of all Spring 2021 NE290 course materials (lecture, assignments, open-source readings) can be accessed via GitHub \url{https://github.com/aaronreichmenberliner/NE290-Spring2021-Nuclear-History-Politics-Futures-}. For additional readings, please contact authors.

\begin{acknowledgments}
We thank the students of NE290 Clara Alivisatos, Preston Awedisean, Michael Bondin, Arnold Eng, Isaac Lipski, Carla McKinley, Austin Mullen, Darren Parkison, Daniel Payne, Chaitanya Peddeti, and Matthew Verlie as well as CAPT. Travis Petzoldt for their patience with our teaching team as we bumbled about Zoom in our first attempt at this new course. We thank the battery of guest lecturers Tim Koeth (University of Maryland), Carl Willis (University of New Mexico), Sarah Schrieber, Mimi Hiebert (University of Maryland), Alex Wellerstein (Stevens Institute of Technology), Anne Harrington (University of Cardiff), Sonja Schmid (Virginia Tech), Marty Pfeiffer (University of New Mexico), Cheryl Rofer (Los Alamos National Laboratory), Ed Geist (RAND), Laura Rockwood (IAEA), and Shirley Johnson (IAEA) for providing their time and expertise to imbuing our course with that \textit{je ne sais quoi}. We thank Dr. Peter Hosemann for his time as the NE290 instructor-of-record and ensuring we did not burn down virtual classroom. We thank our PIs Adam Arkin and Kai Vetter for allowing us to spend our time away from the laboratory. 
\end{acknowledgments}

\bibliography{references}

\cleardoublepage
\end{document}